\documentclass[sigconf]{acmart}

\usepackage{xspace}
\usepackage{enumitem}
\usepackage{algorithmic}
\usepackage[linesnumbered, ruled, vlined]{algorithm2e}
\usepackage{subfigure}
\usepackage{colortbl}
\usepackage{multirow}

\newcommand{\mj}[1]{%
  \begingroup          
  #1
  \endgroup           
}

\newcommand{\ourtool}{SemRef\xspace}
\newcommand{\subsubsubsection}[1]{\noindent\textbf{#1}}
\newcommand{\aaj}{$a2a_{adj}$\xspace}
\newcommand{\ccg}{$c2c_{cvg}$\xspace}

\AtBeginDocument{%
  }

\copyrightyear{2026}
\acmYear{2026}
\setcopyright{cc}
\setcctype{by}
\acmConference[ICSE '26]{2026 IEEE/ACM 48th International Conference on Software Engineering}{April 12--18, 2026}{Rio de Janeiro, Brazil}
\acmBooktitle{2026 IEEE/ACM 48th International Conference on Software Engineering (ICSE '26), April 12--18, 2026, Rio de Janeiro, Brazil}
\acmPrice{}
\acmDOI{10.1145/3744916.3773114}
\acmISBN{979-8-4007-2025-3/2026/04}

\DeclareRobustCommand{\mybox}[2][gray!20]{%
  \par\medskip\noindent
  \colorbox{#1}{%
    \begin{minipage}{\dimexpr\linewidth-2\fboxsep\relax}
      \vspace{2pt}%
      #2%
      \vspace{2pt}%
    \end{minipage}%
  }%
  \par\medskip
}

\begin{document}
\title{Semantic-Enhanced Automatic Refinement of Architecture Recovery Results Using LLMs}

\author{Yiran Zhang}
\email{yiran002@e.ntu.edu.sg}
\affiliation{%
  \institution{Nanyang Technological University}
  \country{Singapore}
}

\author{Chengwei Liu}
\email{chengwei.liu@ntu.edu.sg}
\authornote{Chengwei Liu and Weisong Sun are the corresponding authors.}
\affiliation{%
  \institution{Nanyang Technological University}
  \country{Singapore}
}

\author{Yuqiang Sun}
\email{suny0056@e.ntu.edu.sg}
\affiliation{%
  \institution{Nanyang Technological University}
  \country{Singapore}
}

\author{Zhengzi Xu}
\email{z.xu@imperial.ac.uk}
\affiliation{%
  \institution{Imperial Global Singapore}
  \country{Singapore}
}

\author{Weisong Sun}
\authornotemark[1]
\email{weisong.sun@ntu.edu.sg}
\affiliation{%
  \institution{Nanyang Technological University}
  \country{Singapore}
}

\author{Wenke Li}
\email{lwk949484025@gmail.com}
\affiliation{%
  \institution{Huazhong University of Science and Technology}
  \city{Wuhan}
  \state{Hubei}
  \country{China}
}

\author{Wuxia Jin}
\email{jinwuxia@mail.xjtu.edu.cn}
\affiliation{%
  \institution{Xi'an Jiaotong University}
  \city{Xi'an}
  \state{Shaanxi}
  \country{China}
}

\author{Yang Liu}
\email{yangliu@ntu.edu.sg}
\affiliation{%
  \institution{Nanyang Technological University}
  \country{Singapore}
}


\begin{abstract}
Understanding the architecture is crucial for effectively maintaining and managing large software systems. However, discrepancies often exist between the designed and implemented architectures, which can pose significant risks. To identify these discrepancies, architects need to extract the architecture from the system implementation, which is both time-consuming and error-prone. To simplify this procedure, many automatic architecture recovery techniques have been developed. Yet, their accuracy is often limited. 
Architects must still invest significant effort in refining recovery results to ensure they accurately reflect the implemented architecture.

To reduce such manual effort, we introduce \ourtool, a framework that combines LLMs with dependency analysis to automatically refine architectures recovered by existing architecture recovery tools. 
By leveraging the LLM's semantic understanding capabilities and integrating structural dependencies, \ourtool enhances both the accuracy and the comprehension of recovered architectures.
To evaluate \ourtool, we tested on 9 projects with published ground-truth architectures and 10 state-of-the-art architecture recovery tools. 5 commonly used metrics are adopted to evaluate the effectiveness of \ourtool.
\mj{The results show that \ourtool improves accuracy across various metrics, with normalized gains ranges from 17.72\% to 43.35\%. Specifically, for MoJoFM and \aaj metrics, \ourtool achieves relative improvements of 118.57\% and 100.41\%, respectively.}
Moreover, \ourtool is highly scalable. It maintains stable performance across projects ranging from thousands to trillions of lines of code with the cost scale linearly with project size. Further, we test \ourtool on various LLMs to demonstrate its generalizability across different models.
Beyond improving accuracy, the integration of LLMs enables \ourtool to provide a structured module hierarchy and hierarchical module summaries, which further enhance the comprehensibility of recovered architectures.
\end{abstract}




\maketitle

\section{Introduction}\label{sec_intro}

Due to continuous maintenance activities, software systems often experience a gradual deterioration in architectural clarity, which is known as architectural \textit{drift} and \textit{erosion}~\cite{medvidovic2010software}. This deterioration can result in increased complexity and reduced maintainability of the systems. Consequently, further development may require significantly more effort and resources~\cite{garcia2013obtaining}.

To address these issues, it is essential to timely detect the architectural inconsistencies between the design and its implementation. However, fully understanding the implemented architecture can be highly costly. Recovering the implemented architecture of systems with hundreds of thousands of lines of code may require hundreds of hours from an expert~\cite{garcia2013obtaining}. To alleviate this, various automated software architecture recovery (SAR) tools have been developed to facilitate comprehending the implemented architecture.
For example, ACDC~\cite{tzerpos2000accd}, Bunch~\cite{mancoridis1999bunch}, and FCA~\cite{teymourian_fast_2022} recover the architecture based on the structural dependencies within systems. 
ARC~\cite{garcia2011enhancing} primarily uses semantic analysis, specifically topic modeling of source codes, for recovery.
SADE~\cite{papachristou_software_2019} combines both dependency-based and text-based clustering to analyze the software system.

However, existing SAR tools primarily use structural dependencies and textual similarity to cluster source code files into architectural modules. These methods are coarse-grained and often fail to capture the deeper semantics of software implementations (e.g., utilities and functional modules are designed as modules for various purposes). Such a lack of semantic understanding can lead to coarse-grained results with low accuracy. Therefore, a significant gap persists between these solutions and their real-world applicability. Such a gap necessitates substantial human effort to refine the result of SAR tools to make them suitable for practical use~\cite{lutellier2015comparing}.

Therefore, in this paper, we aim to bridge this gap by integrating enhanced semantic understanding into existing SAR tools. 
With the prevalent adoption of large language models (LLMs) in software analysis tasks, LLMs have demonstrated their powerful capability to understand the semantics of software\mj{~\cite{hou2024large}}. In this paper, we aim to incorporate LLMs for the enhancement of existing SAR tools. 
However, to achieve this, we still face the following challenges:
\textbf{1) Integrating Dependency Analysis with LLM's Semantic Comprehension.} Structural dependencies within a project are vital for comprehending its architecture, as they show the interactions among entities in a system. However, it is challenging for an LLM to directly understand such structural relationships. Integrating dependency information with the LLM's semantic understanding abilities is another critical challenge we face.
\textbf{2) Limited Context Window.} Making architectural decisions requires a high-level understanding of the system. However, projects with architectural concerns are often very large systems. Due to the limited context window, it is impractical to feed the entire project into the LLM. 
How to allow LLMs grasping high-level aspects of large projects beyond their context window presents a significant challenge.

In this paper, we propose \ourtool, a SAR result refinement framework that combines dependency analysis with semantic understanding of LLMs. 
\mj{\ourtool refines the recovered architecture, which contains a list of modules where each module represents a group of source files, in several steps. First, it adjusts module sizes by splitting modules that are too large and merging those that are too small. }
Then, a file-level refinement between modules is done by identifying potential file misplacement with dependency analysis and verifying with LLMs. Next, the flat modules are incrementally built into a hierarchy based on their dependency and semantic similarities. 
Finally, an iterative refinement process is adopted, utilizing the understanding of the project gained from the top-level modules. This process continues until no further abnormalities are detected, thus refining the architecture comprehensively.


To evaluate \ourtool, we collected 9 projects with ground truth architecture and 10 state-of-the-art SAR tools. We applied these tools to the projects, generating a total of 90 recovery results. These results were then refined using \ourtool. The effectiveness of this refinement was evaluated by comparing the architecture similarity, before and after the refinement, with the ground truth. The similarity is then measured by 5 metrics following previous studies~\cite{wen2004effectiveness,le2015empirical,zhang2023software}. Based on the metric result, \ourtool improves the accuracy by 17.72\% to 43.35\% across different metrics, as measured by normalized gain.
Then the effectiveness of each component of our design is examined through an ablation study.
Also, we analyzed the cost of our framework. The result shows that both of them scale linearly with the project size measured by the number of files, which confirms the scalability of \ourtool. Finally, we test \ourtool on different LLMs to demonstrate its generalizability across various models.
In summary, our main contribution includes:

\begin{itemize}[leftmargin=*]
    \item We propose \ourtool, the first LLM-based SAR result refinement framework that integrates dependency analysis with the semantic understanding of LLMs.
    \item We evaluate \ourtool using 10 existing techniques across 9 projects with known ground truth architectures. 5 metrics are utilized to evaluate the improvement. The results show that \ourtool improves the accuracy by 17.72\% to 43.35\% across different metrics, as measured by normalized gain.
    \item We conduct a comprehensive study on \ourtool, confirming its scalability and robustness. The data and code related to our study is publicly available on our website~\cite{oursite}. 

\end{itemize}







\section{Related Work}\label{sec_related}

\subsubsubsection{Software Architecture Recovery (SAR) tools. }
\mj{SAR tools aim to automatically recover software architectures from source code.}
Many of these techniques rely on structural dependencies as their primary source of information. Bunch~\cite{mancoridis1999bunch} clusters source files of software systems by maximizing the modularity quality (MQ) of the dependency graph. WCA~\cite{maqbool2004weighted} is a hierarchical clustering technique often used in software clustering scenarios. LIMBO~\cite{andritsos2004limbo} applies information theory concepts to software clustering. DAGC~\cite{parsa2005new} is similar to Bunch but optimizes its search space. MCA and ECA~\cite{praditwong2010software} incorporate multi-objective optimization into software clustering. CCT~\cite{naseem2013cooperative} is a consensus-based approach utilized in software clustering. Mohammadi et al.~\cite{mohammadi_new_2019} use existing knowledge in the dependency graph to develop a neighborhood tree that guides clustering.
{FCA}~\cite{teymourian_fast_2022} clusters software systems by performing operations on the dependency matrix. 
Unlike these techniques that rely on static dependencies, Xiao et al.~\cite{xiao2005software} demonstrate that dynamic dependencies offer certain advantages.

Apart from dependency-based techniques, many studies use textual information for architecture recovery. ARC~\cite{garcia2011enhancing} recovers architecture using concerns. ZBR~\cite{corazza2011investigating} partitioned textual information into zones to form clusters. Risi et al.~\cite{risi2012using} use LSI to extract textual information and cluster with K-Means. Kargar et al.~\cite{kargar_semantic-based_2017} constructed a semantic dependency graph to replace traditional dependency graphs, later incorporating nominal information~\cite{kargar_multi-programming_2019}. EVOL~\cite{yang_enhancing_2022} improves accuracy through textual outliers filtration and label propagation. Studies show that textual-information-based techniques can outperform dependency-based approaches in certain scenarios~\cite{garcia2013comparative}.
Hybrid techniques utilize both dependency and textual information. Mkaouer et al.~\cite{mkaouer2015many} introduced a many-objective search-based approach using NSGA-III. Chhabra et al.~\cite{chhabra2017improving} extracted features with 24 coupling schemes to optimize clustering. SADE~\cite{cho_software_2019} merges dependency and textual similarity, using text similarity as call graph weights with Louvain clustering. Jalali et al.~\cite{sadat2019multi} developed a multi-objective fitness function treating clustering as a search problem. SARIF~\cite{zhang2023software} combines dependencies, folder structures, and textual information using a dynamic schema.
However, most tools primarily focus on dependency information. Tools incorporating textual information only use embeddings for similarity evaluation, failing to understand semantic relations.

\mj{

\noindent\textbf{LLM for Architecture-Related Tasks.}
Large language models have demonstrated significant potential in various software architecture activities beyond general software engineering tasks. In architectural decision support, LLMs assist architects with design decisions and knowledge access~\cite{dhar2024llmsgeneratearchitecturaldesign,diaz2024helping,maranhao2024prompt,kaplan2024combining}. Several studies generate architectural artifacts from various inputs, such as requirements or informal specifications~\cite{Eisenreich_2024,tagliaferro2025leveraging,soliman2025large}. Recently, focus has shifted toward agent-based automation~\cite{jin2025iredev, lu2025requirements}. Zhang et al.~\cite{zhang2025knowledge} proposed knowledge-driven multi-agent frameworks to automate software architecture design. For architectural quality and comprehension, LLMs help maintain quality through violation resolution and antipattern detection~\cite{rubei2025llm,pandini2025exploratory,mino2024leveraging}. To support high-level system understanding, Sun et al.~\cite{sun2025commenting} explored hierarchical summarization for higher-level code units. Recent work also explores integrating LLMs throughout the architecture lifecycle~\cite{wei2024requirements} and specialized domains~\cite{adnan2025leveraging,hagel2025towards}. Despite these efforts on using LLMs for architecture-related tasks, LLMs' capability in architecture recovery is still rarely explored.
}

\begin{figure*}[tb]
    \centering
    \includegraphics[width=0.95\textwidth]{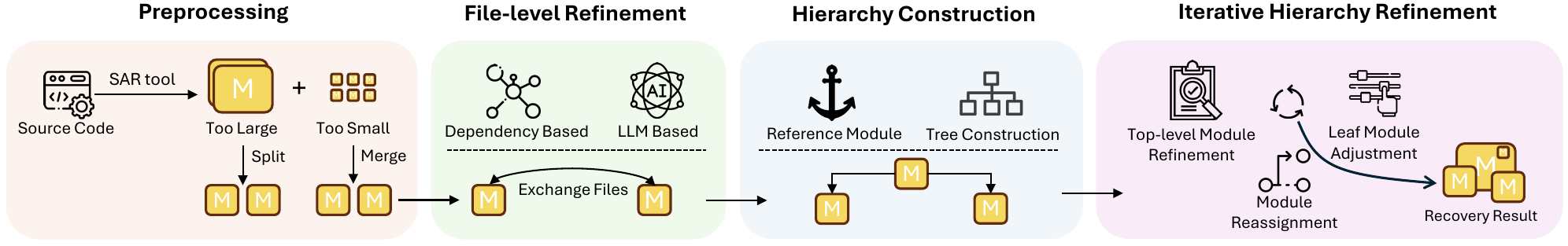}
    \caption{Overview of \ourtool}
    \label{fig_method_overview}
\end{figure*}

\section{Methodology}\label{sec_method}

The overview of our framework is shown in Fig.~\ref{fig_method_overview}.
In the first step, \ourtool performs a coarse-grained refinement of architecture modules whose sizes are overly large or small.
Next, \ourtool detects potential incorrect groupings of files based on dependencies and uses LLM to confirm these potential misgroupings. Following this file-level refinement, \ourtool constructs the module hierarchy based on their structural dependencies and semantic similarities.
After establishing the hierarchy, \ourtool iteratively refines the module structure in a top-down manner. 
The final refined hierarchy serves as the architecture recovery result.


\subsection{Preprocessing of Input Modules}

\mj{For the SAR tools, their recovered architectures are actually a grouping of source files, where each group represents an architectural module. However, the recovered modules often have a large gap between the ground truth modules.}
The most significant issue is that the granularity of the recovered modules often did not match. For example, when we used ACDC to recover the architecture of \textit{ArchStudio 4}, it identified 241 modules. However, 246 out of 2305 (10.67\%) project files were grouped into the same module, while 80 modules contained only one file each. This significant discrepancy made it impractical to represent the project's real module structure: large \mj{recovered modules} might contain several ground truth modules, while single-file \mj{recovered modules} are too small to represent a module effectively. 
\mj{In this preprocessing step, to address this issue, we first divide large modules that contain multiple ground-truth modules into smaller ones. We then merge modules that consist of a single file into larger ones.}


\subsubsection{Splitting Large Modules}\label{sec_split}


\mj{To address the issue of a module containing files from multiple ground-truth modules, we check all modules with more than 2 files to determine if they need to be split.
To mitigate potential hallucination, inspired by the zero-shot chain-of-thought prompting method~\cite{kojima2022large}, we divided the main task into three subtasks.}

In the first sub-task, the LLM is asked to summarize the functionalities of each file independently. Since some of the projects we analyzed are well-known, the LLM's summaries might be influenced by its own knowledge. To mitigate this, we instructed it to avoid considering the project background.
In the second sub-task, we use LLM to summarize the function points of the module based on the file summary from the first task. 
For the third sub-task, we provide the function points and file summaries to determine if the module can be split into several submodules. Moreover, we also provide dependency information to the LLM to enhance its understanding. \mj{Specifically, we identified the weakly connected components within submodules based on their dependencies and provided this information to the LLM for its reference.}
Finally, another prompt is used to format the decision into JSON. Due to the limited space, we refer to our website~\cite{oursite} for the detailed prompt.

\subsubsection{Merge Single File Modules}\label{sec_merge}


In this step, we collect all modules containing a single file and assess whether any of them can be merged into a new module. Similar to the prompt chain design for splitting modules, we first summarize each file with LLM. Next, we evaluate the dependency connectivity among these files \mj{to identify the weakly connected components among the files. Finally, we let LLM decide on the merge based on the summary of the files and their connectivity, where the connectivity is provided to the LLMs as the weakly connected components in the dependency graph.}

\subsection{File-level Refinement}
In the preprocessing step, we refined only the granularity of the modules. Besides the granularity issue, numerous misgroupings at the file level exist, as many SAR tools tend to group completely unrelated files into the same module.


To address these file-level misplacements, we have developed an algorithm that first identifies potential misplacements based on structural dependencies. Then, it verifies the existence of these misplacements using LLM.

\subsubsection{Misplacement Identification via Dependency Check}




Based on the graph, to identify potential misplacements, we assess whether any files have stronger dependencies on another module compared to their own. An intuitive approach is to directly compare the number of file-level dependencies between the file's own module and other modules. However, this method has two main issues.



First, when counting the number of edges, we are essentially measuring the similarity between a file and a module. However, different dependency edges may indicate varying degrees of similarity. For instance, a file used by many others could be a utility file, and dependencies on it might suggest a lower level of similarity. To address this, we normalize the strength of file-level dependencies by the number of dependencies targeting the destination file.
Second, files tend to have more dependencies linked to modules containing more files. Therefore, the dependency strength between a file and a module is normalized by the number of files in the module.
The dependency strength between a file and a module is defined as:

\begin{equation}\label{eq-dep1}
    Dep(f_{src}, {m}_{})=\frac{1}{|{m}_{}|}\cdot \sum_{f\in m_{}}^{}\left(\frac{1}{deg^{-}(f)}\cdot\delta(f_{src}, f)\right)
\end{equation}

where $f_{src}$ and ${m}$ represent  the file and module, respectively. $|{m}_{}|$ is the number of files in $m$, $deg^{-}(f)$ is the in-degree of $f$ (i.e., number of files that depend on $f$), and $\delta(f_{src}, f)$ is 1 if $f_{src}$ depends on $f$; otherwise it is 0.


For each file in the project, we calculate its dependency weight to all modules. If any module exhibits a higher dependency weight than the file's own module, the file is marked to be potentially misplaced. In the next step, we will determine whether the file should remain in its current module or be moved to the module with the stronger dependency.

\subsubsection{LLM Verification on Potential Misplacements}


In this step, all potential misplacements identified in the last step are verified using the LLM to determine if they are false positives. We iterate over modules with potential misplacements one by one. For each module, to prevent the LLM from being influenced by the original grouping, we first remove all potentially misplaced files from the module. Then, we place the files back one by one.




\mj{For each file, we first sort its candidate modules from the previous step based on their dependency strength. We then evaluate each candidate module in order to determine if the file should be moved. For this evaluation, we provide the LLM with information about both the file's original module and the candidate module. This information includes a list of files within each module and their corresponding summaries.}
Then, we let LLM decide which of the two modules the file should belong to. If the file is moved to a new module, we will temporarily move the file to the candidate module and ask LLM if the file has a closer semantic relation with the new module than the original one. If so, the file will be moved. Otherwise, we deny the movement and go on to check the following candidate modules. To prevent potential inaccuracies due to LLM hallucinations, if a file is examined more than three times and consistently remains in its original module, it will be deemed correctly organized and will not be further checked. 
This algorithm will end after all potential misplacements are checked.

\subsection{Hierarchy Construction}
In this step, we aim to organize the modules into hierarchy. The motivation for this step is that the flat view of modules is not good for LLM to understand the project's scope. On one hand, when the number of modules is large, it may exceed LLM's context window. On the other hand, the leaf modules can have different abstraction level, input all the modules together to LLM could caused biased understanding. By organizing the modules into hierarchy, the LLM can understand the entire project hierarchically from bottom up, which can address these two issues.
\mj{We use a two-step algorithm to construct the module hierarchy. First, we identify a set of reference modules with clear responsibilities to serve as the foundation. Second, we incrementally place the remaining modules into the hierarchy based on their relationship to these reference modules.}

\subsubsection{Add Reference Modules}




The first step is to select the reference modules.  As a reference module, we want each reference module to cover a small but cohesive set of responsibilities within the system. Additionally, each selected reference module should have a distinct responsibility with no overlap with the others.

To identify cohesive modules, we measure their dependency cohesion as the percentage of internal dependencies among all dependencies related to the module:

\begin{equation}\label{eq-cohesion}
    Cohesion(m_{src}) = \frac{\sum_{f\in m_{src}}^{}Dep(f,m_{src})}{\sum_{m\in project}^{}\sum_{f\in m_{src}}^{}Dep(f,m)}
\end{equation}

where $Dep(f,m)$ is defined in Eq.~\ref{eq-dep1}. 
Next, we select the set of reference modules based on cohesion. First, the module with the highest cohesion is chosen as the initial reference module. Then, we traverse the remaining modules, sorted in descending order of cohesion. To prevent overlap, a module is added to the reference set only if it has no dependency relationship with any module already in the set. This selection process continues until all modules have been examined.

\subsubsection{Construct Tree Structure}


After selecting the reference modules, we first establish a single-level module hierarchy by creating a root and placing all reference modules under it. Next, we integrate the remaining modules into the hierarchy.
For each module, we instruct the LLM to determine its placement by identifying the most semantically related modules in the existing hierarchy. To assist in this process, the LLM is provided with the following information: 1) a summary of the module to be added along with the names of its files, 2) the current module hierarchy, 3) modules within the hierarchy that have dependencies on the module to be added, and 4) summaries of the top-level modules. 
\mj{Initially, the LLM receives only the summaries of top-level modules. If more information is needed to make a decision, it can request summaries of other modules. This retrieval is fully automated. The LLM uses the module path to request a specific summary, and the system then finds the corresponding module and provides the summary.}

To clarify the hierarchy, we adopt a different organizational policy than typical folder structures, which mix files and subfolders within the same directory. In our approach, a module can contain either files or submodules exclusively. In other words, only leaf modules can contain files, while branch modules (i.e., non-leaf modules) can only contain submodules. To maintain this structure, when placing a module, we allow the LLM to choose only between two actions: 1) positioning the module under an existing branch module (including the root) or 2) creating a new branch module to accommodate this module and other similar modules.


After a module is added, an iterative summarization process is triggered for the hierarchy affected by the new module's addition. For leaf modules, the summary is derived from the files they contain and their respective summaries. For branch modules, the summary is generated based on the summaries of their submodules.
\mj{As the hierarchy could be inaccurate, the summary here could be wrong. We will address this issue in the following hierarchy refinement process.
The module placement continues until all modules have been incorporated into the hierarchy.}




\subsection{Hierarchy Refinement}

After constructing the initial module hierarchy, the final task is to refine it. This initial hierarchy often has many issues: 1) the top-level modules (i.e. the module directly under root) may not accurately reflect the system's key components, 2) modules may be placed under a parent with which they have a weak relationship, and 3) leaf modules may still contain misgrouped files.



To address these concerns, we designed a top-down approach to iteratively refine the hierarchy. First, we refine the top-level modules to ensure each has a clear boundary and no overlapping concepts. Next, we systematically review all remaining modules to determine whether they are too vague or placed under an incorrect parent. Finally, we revisit the content of the leaf modules to assess whether further refinement is needed.

Unlike the previous steps, thanks to the hierarchy, the refinement process can have a more comprehensive understanding of the project. However, this understanding is primarily derived from the hierarchical summary of the system's structure, while the structure itself is shaped by this understanding. To resolve this interdependence, we iterate through the three refinement steps until both the hierarchy and the system's understanding become stable.




\subsubsection{Refine Top-level Modules}

The first step is to refine the top-level modules. To improve cohesion and reduce coupling, each module should have a clear responsibility boundary, and ensure that no module overlaps in responsibility with the others.
To achieve this, we first refine top-level modules by identifying and splitting those that cover overly broad responsibilities. Next, we identify groups of modules with overlapping responsibilities and resolve these overlaps to create a well-structured hierarchy.

\subsubsubsection{Splitting Vague Modules.}
To determine whether a top-level module is too vague or covers an overly broad scope without a clear boundary, we provide the LLM with the following information: 1) a summary of the entire system to help the LLM understand its overall scope, 2) a summary of the module under evaluation along with its submodules to clarify its specific scope, and 3) the cohesion scores of both the module under evaluation and its submodules to assess their cohesiveness from a dependency perspective.

\mj{Then we use LLM to determine if a module should be split. The decision is made based on the LLM's own judgment on the semantic cohesion of the module based on the provided information. This is an iterative process that begins by adding all top-level modules to an evaluation queue. When a module is evaluated, the LLM decides if it should be split. If a module is split, its new submodules are added to the queue for later examination. The process repeats until the queue is empty and all modules have been checked.}

\subsubsubsection{Resolve Overlapping Modules.} 
After splitting, we identify potential overlaps among the modules. In this step, we evaluate the top-level modules one by one. For each module, we provide the LLM with the following information: 1) a summary of the module under evaluation along with summaries of all other top-level modules and 2) the dependency strength between the module under evaluation and each remaining top-level module, measured as follows:

\begin{equation}\label{eq-dep-module}
    Dep_m(m_{src}, {m}_{dst})=\sum_{f\in m_{src}}^{}Dep(f,m_{dst})
\end{equation}
where $Dep(f,m_{dst})$ is defined in Eq.~\ref{eq-dep1}. 
To identify overlaps, we instruct the LLM to: 1) understand the scope of the module under evaluation and 2) compare it with each remaining module one by one to determine whether they share overlapping responsibilities.

If potential overlaps are identified, we instruct the LLM to resolve them. To do this, we provide the LLM with information about the identified overlapping modules and their submodules, allowing it to freely reorganize them to eliminate overlapping.
\mj{After modifying the hierarchy, we evaluate if the change should be kept. First, we calculate the total cohesion of the top-level modules using the sum of their scores from $Eq.~\ref{eq-cohesion}$. If this total score increases, which means the dependency cohesion aligns with the semantic cohesion as judged by LLM,  we accept the modification. If the score does not increase, we then ask the LLM to compare the responsibility cohesion of the original and the modified hierarchies. In this second case, we apply the changes only if the LLM confirms that the new hierarchy is better.}


\subsubsection{Module Reassignment} 

After adjusting the top-level modules, our goal is to ensure that the remaining modules are correctly placed within the hierarchy. To achieve this, we perform a pre-order traversal of the entire hierarchy.
For each module (except top-level ones), we ask the LLM whether its responsibilities align well with its current parent. If the LLM determines that the module is not well aligned, we ask it to select a more appropriate placement. 
Besides this candidate provided by LLM, we also include the top three modules with the strongest dependency strength, as measured by Eq.~\ref{eq-dep-module}, in relation to the module under evaluation.
In addition to the candidate provided by the LLM, we also include the modules with the top three strongest dependency strength.


Then, we use a separate prompt to let the LLM determine where the module should be placed. To mitigate potential hallucinations, we check whether the module has a stronger dependency with the new parent or the original one. If it does, we accept the movement. If not, we ask the LLM to double-check its decision by providing detailed information about the two potential parents, including 1) a summary of each parent module, 2) the sub-hierarchy of the parent along with summaries of its submodules, and 3) the dependency details between the parent module and the module under evaluation. If this double-check confirms the previous decision, we accept the change and move the module to the new parent, even if it contradicts the dependency strength.

\subsubsection{Refine Leaf Modules}
After refining the misplaced modules, we finally revisit the quality of the leaf modules. Unlike the steps before hierarchy construction, we can now provide the LLM with the project’s context to improve the accuracy of the refinement. For each prompt used in leaf module refinement, we include the module hierarchy, the project summary, and the top-level module summaries to help the LLM understand the project organization.

The refinement begins with identifying and relocating misplaced files. Similar to module movement, we use the LLM to determine which files are weakly related to their current module by providing it with the module summary and summaries of each file within the module. For each potentially misplaced file, we use a separate prompt to evaluate whether it should be moved elsewhere. The alternative placement candidates come from two sources: 1) modules that the LLM identifies as semantically related and 2) the top three modules with the strongest dependency strength with the file, as measured by Eq.~\ref{eq-dep1}. The LLM then determines the final placement of the file. A double-check is triggered if the file is suggested to be moved to a module with weak dependency to it.



After relocating unrelated files, we use the LLM to determine whether the module should be split or merged with other modules. To do this, we first check whether the module covers too many responsibilities of its parent module compared to its sibling modules. If so, a module split process, as detailed in Sec.~\ref{sec_split}, is triggered to assess whether splitting is necessary. If the module is decided to be split, we perform a double-check by presenting the LLM with the module organization before and after the split. The LLM then evaluates whether one organization is significantly better than the other. The split is executed only if the LLM determines that the post-split organization is significantly better.

Finally, we assess whether the module should be merged. Similar to the split process, first, the LLM determines whether the module covers too few responsibilities of its parent module. If so, we let the LLM decide whether the module should be merged into one of its sibling modules. If a merge is suggested, it is executed only if the LLM confirms that the new hierarchy after the merge is significantly better than the previous one.


\subsubsection{Update of Summarization} 
After each modification during the three refinement steps, a hierarchical resummarization is triggered. First, modules with direct content changes are resummarized. To minimize overhead and instability, before each resummarization, we first let the LLM determine whether the current summary remains accurate. The summary is updated only if the LLM determines that it is no longer accurate. If a module’s summary is updated, a similar resummarization check is performed on its parent, continuing until the LLM decides that no further updates are needed.

\subsubsection{Iterate Until Stable}

The three hierarchy refinement steps will iterate until they stabilize. As mentioned earlier, this iteration is necessary because updating the hierarchy changes its summaries, which may influence the LLM’s decisions on further refinements. To prevent unnecessary re-evaluations, a check is only re-triggered if the content provided to the LLM has changed.

More specifically, for top module refinement, if the project summary changed in the last round, the entire process is repeated. Otherwise, only the top modules with updated summaries from the last round are checked. For module reassignment, a module is skipped if its siblings and parent have not changed in the last round. Similarly, for leaf module refinement, a module is rechecked only if it, its parent, or its siblings have changed since the last round. The iteration continues until no modifications are made in a full round.



\section{Evaluation}\label{sec_eval}

{
\setlength{\tabcolsep}{1.85pt}
\begin{table}[tb]
    \caption{Information of projects in evaluation}
    \label{tab_projects}\footnotesize
    \centering
    \begin{tabular}{l|llllll}
        \hline 
        Name & Version & Lang. & NLOC & \#File & \#Module & Domain\\
        \hline 
        ArchStudio & 4  & Java & 238K & 2305 & 57 & IDE\\
        Bash & 4.2  & C & 115K & 405 & 14 & Shell\\
        Chromium & svn-171054 & {C++} & 11.7M & 46,498 & 67 & Browser \\
        HDC & 46ff87  & {C++} & 25.7K & 207 & 11 & {Camera Interface} \\
        HDF & 0e196f  & {C} & 153K & 1,051 & 15 & {Driver Subsystem}\\
        Hadoop & 0.19.0  & Java & 225K & 1,703 & 67 & Dist. Computing \\
        ITK & 4.5.2 & C++ & 1.23M & 8,504 & 11 & Image Processing\\
        Libxml2  & 2.4.22 & C & 81.1K & 82 & 17 & {XML Parser} \\
        OODT & 0.2  & Java & 81.4K & 892 & 216 & Data Management\\
        \hline 
    \end{tabular}
\end{table}
}

In this section, we evaluate \ourtool by answering the following research questions:

\begin{itemize}[left=0.5em]
    \item \textbf{RQ1 (Effectiveness Analysis):} How effective is \ourtool in increasing the accuracy of the SAR tools?
    \item \textbf{RQ2 (Ablation Study):} 
    How \mj{does} each part of SemRef contribute to \mj{increased} accuracy?
    \item \textbf{RQ3 (Cost Analysis):} 
    How many tokens \mj{does} SemRef require for the architecture refinement of projects of different sizes?
    \item \textbf{RQ4 (Model Comparison):} How \mj{does} using different LLMs affect the \mj{effectiveness} of \ourtool?
\end{itemize}

{
\setlength{\tabcolsep}{2.85pt}
\begin{table*}[tb]
    \centering
    \caption{The metric scores for input architecture and architecture refined by SemRef. All metric values are timed for 100. (Abbreviations: M-MoJoFM, A-a2a, R-ARI, J-\aaj, C-\ccg\mj{, I-Input, O-Output, D-Diff in PP.})} \footnotesize
    \label{tab_metrics_result}
    \scalebox{0.8}{
    \begin{tabular}{lc|ccccc|ccccc|ccccc|ccccc|ccccc|ccccc|ccccc|ccccc|ccccc}
        \hline
         && \multicolumn{5}{c|}{ArchStudio} & \multicolumn{5}{c|}{Bash} & \multicolumn{5}{c|}{Chromium} & \multicolumn{5}{c|}{HDC} & \multicolumn{5}{c|}{HDF} & \multicolumn{5}{c|}{Hadoop} & \multicolumn{5}{c|}{ITK} & \multicolumn{5}{c|}{Libxml2} & \multicolumn{5}{c}{OODT}\\ 
        && {\scriptsize M} & {\scriptsize A} & {\scriptsize R} & {\scriptsize J} & {\scriptsize C} & {\scriptsize M} & {\scriptsize A} & {\scriptsize R} & {\scriptsize J} & {\scriptsize C} & {\scriptsize M} & {\scriptsize A} & {\scriptsize R} & {\scriptsize J} & {\scriptsize C} & {\scriptsize M} & {\scriptsize A} & {\scriptsize R} & {\scriptsize J} & {\scriptsize C} & {\scriptsize M} & {\scriptsize A} & {\scriptsize R} & {\scriptsize J} & {\scriptsize C} & {\scriptsize M} & {\scriptsize A} & {\scriptsize R} & {\scriptsize J} & {\scriptsize C} & {\scriptsize M} & {\scriptsize A} & {\scriptsize R} & {\scriptsize J} & {\scriptsize C} & {\scriptsize M} & {\scriptsize A} & {\scriptsize R} & {\scriptsize J} & {\scriptsize C} & {\scriptsize M} & {\scriptsize A} & {\scriptsize R} & {\scriptsize J} & {\scriptsize C}\\ \hline\hline
        \multirow{3}{*}{ACDC} & I & 73& 86& 20& 50& 24& 53& 81& 12& 40&  4& 48& 82&  7& 37&  1& 49& 82& 11& 31&  0& 38& 79& 13& 38&  0& 46& 82& 13& 33&  5& 40& 73&  2& 25&  0& 34& 83& 13& 40& 25& 44& 85& 24& 45& 17\\
        & O & 83& 88& 44& 64& 40& 69& 86& 29& 53&  9& 71& 89& 38& 56& 14& 79& 87& 32& 57& 29& 65& 88& 36& 56& 12& 68& 86& 33& 49& 14& 68& 85&  8& 42&  9& 56& 87& 32& 52& 31& 62& 89& 45& 59& 36\\
        & D &\cellcolor{black!20}{10}&\cellcolor{black!20}{2}&\cellcolor{black!20}{23}&\cellcolor{black!20}{14}&\cellcolor{black!20}{16}&\cellcolor{black!20}{16}&\cellcolor{black!20}{5}&\cellcolor{black!20}{17}&\cellcolor{black!20}{13}&\cellcolor{black!20}{5}&\cellcolor{black!20}{23}&\cellcolor{black!20}{6}&\cellcolor{black!20}{31}&\cellcolor{black!20}{19}&\cellcolor{black!20}{14}&\cellcolor{black!20}{30}&\cellcolor{black!20}{5}&\cellcolor{black!20}{21}&\cellcolor{black!20}{25}&\cellcolor{black!20}{29}&\cellcolor{black!20}{27}&\cellcolor{black!20}{9}&\cellcolor{black!20}{23}&\cellcolor{black!20}{18}&\cellcolor{black!20}{12}&\cellcolor{black!20}{22}&\cellcolor{black!20}{4}&\cellcolor{black!20}{21}&\cellcolor{black!20}{16}&\cellcolor{black!20}{9}&\cellcolor{black!20}{28}&\cellcolor{black!20}{12}&\cellcolor{black!20}{6}&\cellcolor{black!20}{17}&\cellcolor{black!20}{9}&\cellcolor{black!20}{22}&\cellcolor{black!20}{4}&\cellcolor{black!20}{19}&\cellcolor{black!20}{12}&\cellcolor{black!20}{6}&\cellcolor{black!20}{18}&\cellcolor{black!20}{4}&\cellcolor{black!20}{21}&\cellcolor{black!20}{13}&\cellcolor{black!20}{20}\\ \hline
        \multirow{3}{*}{ARC} & I & 35& 83&  9& 39&  0& 26& 79& -1& 10&  0& 32& 83&  0& 32&  0& 30& 79&  1& 14&  0& 17& 80& -1& 13&  0& 17& 79&  1& 25&  7& 76& 77&  0& 32&  0& 27& 83&  6& 31&  5&  8& 75&  0& 24&  5\\
        & O & 70& 88& 31& 56& 23& 57& 87& 21& 53&  9& 69& 90& 40& 58& 17& 62& 84& 26& 43& 12& 68& 87& 43& 59& 18& 63& 88& 56& 56& 21& 82& 85& 21& 48& 13& 75& 91& 45& 69& 40& 61& 89& 48& 56& 30\\
        & D &\cellcolor{black!20}{35}&\cellcolor{black!20}{5}&\cellcolor{black!20}{22}&\cellcolor{black!20}{16}&\cellcolor{black!20}{23}&\cellcolor{black!20}{31}&\cellcolor{black!20}{8}&\cellcolor{black!20}{22}&\cellcolor{black!20}{43}&\cellcolor{black!20}{9}&\cellcolor{black!20}{37}&\cellcolor{black!20}{7}&\cellcolor{black!20}{40}&\cellcolor{black!20}{27}&\cellcolor{black!20}{17}&\cellcolor{black!20}{32}&\cellcolor{black!20}{6}&\cellcolor{black!20}{25}&\cellcolor{black!20}{29}&\cellcolor{black!20}{12}&\cellcolor{black!20}{51}&\cellcolor{black!20}{7}&\cellcolor{black!20}{44}&\cellcolor{black!20}{46}&\cellcolor{black!20}{18}&\cellcolor{black!20}{46}&\cellcolor{black!20}{9}&\cellcolor{black!20}{56}&\cellcolor{black!20}{32}&\cellcolor{black!20}{14}&\cellcolor{black!20}{6}&\cellcolor{black!20}{8}&\cellcolor{black!20}{21}&\cellcolor{black!20}{16}&\cellcolor{black!20}{13}&\cellcolor{black!20}{48}&\cellcolor{black!20}{8}&\cellcolor{black!20}{40}&\cellcolor{black!20}{38}&\cellcolor{black!20}{35}&\cellcolor{black!20}{54}&\cellcolor{black!20}{14}&\cellcolor{black!20}{48}&\cellcolor{black!20}{33}&\cellcolor{black!20}{25}\\ \hline
        \multirow{3}{*}{Bunch\_N} & I & 39& 82& 14& 34&  6& 41& 83& 10& 29&  0& 57& 73&  0& 20&  0& 55& 86& 24& 37&  0& 42& 84& 11& 35&  0& 34& 82& 12& 31&  0& 40& 80&  1& 15&  0& 22& 81&  9& 36&  0& 14& 77&  6& 29&  0\\
        & O & 72& 89& 42& 55& 26& 72& 87& 33& 53&  6& 69& 85& 26& 48& 12& 68& 89& 30& 49& 18& 65& 86& 29& 52& 17& 62& 87& 46& 47& 17& 64& 87& 24& 49& 12& 51& 89& 81& 51& 38& 46& 86& 24& 48& 24\\
        & D &\cellcolor{black!20}{33}&\cellcolor{black!20}{7}&\cellcolor{black!20}{28}&\cellcolor{black!20}{21}&\cellcolor{black!20}{19}&\cellcolor{black!20}{31}&\cellcolor{black!20}{3}&\cellcolor{black!20}{23}&\cellcolor{black!20}{24}&\cellcolor{black!20}{6}&\cellcolor{black!20}{11}&\cellcolor{black!20}{11}&\cellcolor{black!20}{26}&\cellcolor{black!20}{28}&\cellcolor{black!20}{12}&\cellcolor{black!20}{13}&\cellcolor{black!20}{4}&\cellcolor{black!20}{6}&\cellcolor{black!20}{13}&\cellcolor{black!20}{18}&\cellcolor{black!20}{23}&\cellcolor{black!20}{2}&\cellcolor{black!20}{18}&\cellcolor{black!20}{17}&\cellcolor{black!20}{17}&\cellcolor{black!20}{28}&\cellcolor{black!20}{5}&\cellcolor{black!20}{34}&\cellcolor{black!20}{16}&\cellcolor{black!20}{17}&\cellcolor{black!20}{25}&\cellcolor{black!20}{7}&\cellcolor{black!20}{23}&\cellcolor{black!20}{35}&\cellcolor{black!20}{12}&\cellcolor{black!20}{29}&\cellcolor{black!20}{8}&\cellcolor{black!20}{73}&\cellcolor{black!20}{15}&\cellcolor{black!20}{38}&\cellcolor{black!20}{32}&\cellcolor{black!20}{9}&\cellcolor{black!20}{18}&\cellcolor{black!20}{19}&\cellcolor{black!20}{24}\\ \hline
        \multirow{3}{*}{Bunch\_S} & I & 62& 85& 22& 41&  7& 51& 84& 14& 35&  4& 27& 74&  0& 12&  0& 65& 88& 34& 48&  0& 40& 86& 17& 35&  0& 34& 81& 15& 33&  0& 54& 78&  4& 14&  0& 38& 86& 20& 47& 11& 14& 77&  6& 29&  0\\
        & O & 73& 89& 39& 57& 38& 68& 89& 35& 61& 17& 72& 84& 38& 48& 15& 77& 90& 42& 57& 19& 67& 89& 38& 61& 17& 56& 87& 39& 50& 22& 66& 87& 28& 47& 15& 61& 90& 29& 59& 24& 48& 86& 30& 50& 30\\
        & D &\cellcolor{black!20}{11}&\cellcolor{black!20}{4}&\cellcolor{black!20}{18}&\cellcolor{black!20}{16}&\cellcolor{black!20}{31}&\cellcolor{black!20}{17}&\cellcolor{black!20}{6}&\cellcolor{black!20}{21}&\cellcolor{black!20}{25}&\cellcolor{black!20}{13}&\cellcolor{black!20}{45}&\cellcolor{black!20}{10}&\cellcolor{black!20}{38}&\cellcolor{black!20}{36}&\cellcolor{black!20}{15}&\cellcolor{black!20}{13}&\cellcolor{black!20}{2}&\cellcolor{black!20}{8}&\cellcolor{black!20}{9}&\cellcolor{black!20}{19}&\cellcolor{black!20}{27}&\cellcolor{black!20}{3}&\cellcolor{black!20}{20}&\cellcolor{black!20}{26}&\cellcolor{black!20}{17}&\cellcolor{black!20}{22}&\cellcolor{black!20}{6}&\cellcolor{black!20}{24}&\cellcolor{black!20}{17}&\cellcolor{black!20}{22}&\cellcolor{black!20}{12}&\cellcolor{black!20}{9}&\cellcolor{black!20}{24}&\cellcolor{black!20}{33}&\cellcolor{black!20}{15}&\cellcolor{black!20}{22}&\cellcolor{black!20}{4}&\cellcolor{black!20}{9}&\cellcolor{black!20}{12}&\cellcolor{black!20}{13}&\cellcolor{black!20}{34}&\cellcolor{black!20}{9}&\cellcolor{black!20}{23}&\cellcolor{black!20}{21}&\cellcolor{black!20}{30}\\ \hline
        \multirow{3}{*}{FCA} & I & 60& 82& 10& 43&  9& 45& 81& 16& 44&  2& 58& 76&  3& 29&  0& 53& 78& 10& 35&  0& 32& 78& 15& 41&  0& 49& 82& 11& 38&  7& 44& 73&  1& 25&  0& 37& 84&  8& 37& 10& 50& 86& 23& 44& 14\\
        & O & 77& 89& 40& 60& 39& 70& 87& 33& 57& 12& 69& 84& 23& 47& 16& 67& 85& 31& 52& 10& 76& 89& 47& 62& 16& 60& 87& 46& 51& 31& 63& 83& 29& 48&  9& 64& 88& 30& 61& 36& 53& 87& 32& 49& 28\\
        & D &\cellcolor{black!20}{17}&\cellcolor{black!20}{7}&\cellcolor{black!20}{30}&\cellcolor{black!20}{18}&\cellcolor{black!20}{30}&\cellcolor{black!20}{26}&\cellcolor{black!20}{6}&\cellcolor{black!20}{17}&\cellcolor{black!20}{14}&\cellcolor{black!20}{9}&\cellcolor{black!20}{12}&\cellcolor{black!20}{8}&\cellcolor{black!20}{20}&\cellcolor{black!20}{18}&\cellcolor{black!20}{16}&\cellcolor{black!20}{14}&\cellcolor{black!20}{7}&\cellcolor{black!20}{21}&\cellcolor{black!20}{17}&\cellcolor{black!20}{10}&\cellcolor{black!20}{44}&\cellcolor{black!20}{11}&\cellcolor{black!20}{32}&\cellcolor{black!20}{21}&\cellcolor{black!20}{16}&\cellcolor{black!20}{11}&\cellcolor{black!20}{5}&\cellcolor{black!20}{35}&\cellcolor{black!20}{13}&\cellcolor{black!20}{25}&\cellcolor{black!20}{19}&\cellcolor{black!20}{10}&\cellcolor{black!20}{28}&\cellcolor{black!20}{23}&\cellcolor{black!20}{9}&\cellcolor{black!20}{27}&\cellcolor{black!20}{4}&\cellcolor{black!20}{22}&\cellcolor{black!20}{24}&\cellcolor{black!20}{27}&\cellcolor{black!20}{4}&\cellcolor{black!20}{1}&\cellcolor{black!20}{9}&\cellcolor{black!20}{5}&\cellcolor{black!20}{14}\\ \hline
        \multirow{3}{*}{LIMBO} & I & 23& 79& -0& 11&  0& 27& 77&  0& 17&  0& 30& 76&  0&  3&  0& 23& 77&  0& 22&  0& 15& 76& -1& 25&  0& 14& 79&  0& 13&  0& 49& 76&  0&  3&  0& 15& 80& -0& 38&  2&  7& 76& -0& 21&  0\\
        & O & 68& 88& 43& 56& 30& 65& 86& 27& 54& 11& 58& 85& 26& 32&  6& 74& 89& 48& 64& 21& 70& 92& 46& 67& 40& 51& 86& 38& 42&  9& 69& 82& 19& 32&  9& 44& 88& 29& 49& 19& 35& 84& 12& 34&  5\\
        & D &\cellcolor{black!20}{45}&\cellcolor{black!20}{9}&\cellcolor{black!20}{44}&\cellcolor{black!20}{45}&\cellcolor{black!20}{30}&\cellcolor{black!20}{38}&\cellcolor{black!20}{9}&\cellcolor{black!20}{27}&\cellcolor{black!20}{37}&\cellcolor{black!20}{11}&\cellcolor{black!20}{28}&\cellcolor{black!20}{9}&\cellcolor{black!20}{26}&\cellcolor{black!20}{29}&\cellcolor{black!20}{6}&\cellcolor{black!20}{51}&\cellcolor{black!20}{12}&\cellcolor{black!20}{48}&\cellcolor{black!20}{42}&\cellcolor{black!20}{21}&\cellcolor{black!20}{56}&\cellcolor{black!20}{16}&\cellcolor{black!20}{47}&\cellcolor{black!20}{42}&\cellcolor{black!20}{40}&\cellcolor{black!20}{38}&\cellcolor{black!20}{7}&\cellcolor{black!20}{38}&\cellcolor{black!20}{29}&\cellcolor{black!20}{9}&\cellcolor{black!20}{20}&\cellcolor{black!20}{6}&\cellcolor{black!20}{19}&\cellcolor{black!20}{29}&\cellcolor{black!20}{9}&\cellcolor{black!20}{29}&\cellcolor{black!20}{7}&\cellcolor{black!20}{30}&\cellcolor{black!20}{11}&\cellcolor{black!20}{17}&\cellcolor{black!20}{28}&\cellcolor{black!20}{7}&\cellcolor{black!20}{12}&\cellcolor{black!20}{13}&\cellcolor{black!20}{5}\\ \hline
        \multirow{3}{*}{SADE} & I & 42& 83& 11& 41&  0& 52& 87& 20& 46& 17& 58& 87& 26& 48& 24& 72& 88& 41& 52&  8& 49& 88& 30& 44& 22& 41& 83& 23& 44& 22& 60& 81&  9& 24&  0& 34& 84& 22& 44& 17& 21& 78& 13& 34& 17\\
        & O & 80& 92& 61& 70& 44& 79& 89& 40& 60& 19& 71& 89& 39& 53& 31& 79& 92& 44& 57& 16& 67& 90& 42& 54& 25& 60& 86& 39& 53& 23& 70& 87& 25& 36& 12& 57& 90& 33& 55& 29& 60& 89& 52& 63& 43\\
        & D &\cellcolor{black!20}{39}&\cellcolor{black!20}{9}&\cellcolor{black!20}{50}&\cellcolor{black!20}{28}&\cellcolor{black!20}{44}&\cellcolor{black!20}{27}&\cellcolor{black!20}{2}&\cellcolor{black!20}{19}&\cellcolor{black!20}{14}&\cellcolor{black!20}{3}&\cellcolor{black!20}{12}&\cellcolor{black!20}{2}&\cellcolor{black!20}{13}&\cellcolor{black!20}{5}&\cellcolor{black!20}{7}&\cellcolor{black!20}{7}&\cellcolor{black!20}{3}&\cellcolor{black!20}{3}&\cellcolor{black!20}{5}&\cellcolor{black!20}{8}&\cellcolor{black!20}{18}&\cellcolor{black!20}{2}&\cellcolor{black!20}{12}&\cellcolor{black!20}{10}&\cellcolor{black!20}{3}&\cellcolor{black!20}{19}&\cellcolor{black!20}{3}&\cellcolor{black!20}{16}&\cellcolor{black!20}{10}&\cellcolor{black!20}{1}&\cellcolor{black!20}{10}&\cellcolor{black!20}{5}&\cellcolor{black!20}{16}&\cellcolor{black!20}{11}&\cellcolor{black!20}{12}&\cellcolor{black!20}{22}&\cellcolor{black!20}{6}&\cellcolor{black!20}{11}&\cellcolor{black!20}{11}&\cellcolor{black!20}{13}&\cellcolor{black!20}{38}&\cellcolor{black!20}{11}&\cellcolor{black!20}{39}&\cellcolor{black!20}{28}&\cellcolor{black!20}{25}\\ \hline
        \multirow{3}{*}{SARIF} & I & 67& 87& 33& 51& 42& 77& 90& 51& 67& 14& 63& 87& 32& 49& 17& 89& 88& 51& 54& 16& 59& 87& 35& 50& 18& 54& 87& 52& 56& 22& 68& 84& 21& 35&  0& 53& 90& 38& 58& 23& 24& 79& 17& 37& 13\\
        & O & 84& 92& 63& 70& 46& 82& 91& 62& 72& 26& 79& 90& 47& 57& 24& 92& 92& 62& 63& 33& 68& 90& 44& 58& 29& 68& 92& 70& 66& 41& 89& 88& 42& 49& 19& 74& 92& 46& 69& 41& 61& 89& 53& 63& 51\\
        & D &\cellcolor{black!20}{17}&\cellcolor{black!20}{5}&\cellcolor{black!20}{29}&\cellcolor{black!20}{19}&\cellcolor{black!20}{4}&\cellcolor{black!20}{6}&\cellcolor{black!20}{1}&\cellcolor{black!20}{11}&\cellcolor{black!20}{5}&\cellcolor{black!20}{12}&\cellcolor{black!20}{16}&\cellcolor{black!20}{2}&\cellcolor{black!20}{15}&\cellcolor{black!20}{7}&\cellcolor{black!20}{6}&\cellcolor{black!20}{3}&\cellcolor{black!20}{4}&\cellcolor{black!20}{11}&\cellcolor{black!20}{10}&\cellcolor{black!20}{18}&\cellcolor{black!20}{8}&\cellcolor{black!20}{2}&\cellcolor{black!20}{8}&\cellcolor{black!20}{8}&\cellcolor{black!20}{11}&\cellcolor{black!20}{14}&\cellcolor{black!20}{4}&\cellcolor{black!20}{18}&\cellcolor{black!20}{10}&\cellcolor{black!20}{18}&\cellcolor{black!20}{22}&\cellcolor{black!20}{4}&\cellcolor{black!20}{21}&\cellcolor{black!20}{14}&\cellcolor{black!20}{19}&\cellcolor{black!20}{20}&\cellcolor{black!20}{2}&\cellcolor{black!20}{8}&\cellcolor{black!20}{10}&\cellcolor{black!20}{18}&\cellcolor{black!20}{37}&\cellcolor{black!20}{10}&\cellcolor{black!20}{37}&\cellcolor{black!20}{25}&\cellcolor{black!20}{38}\\ \hline
        \multirow{3}{*}{WCA\_UE} & I & 31& 82&  3& 34&  0& 24& 80& -1& 31&  0& 31& 82&  2& 29&  0& 41& 80& 11& 33&  0& 35& 82& 12& 37&  4& 13& 78& -1& 17&  0& 49& 85& -7& 42&  0& 29& 82&  6& 44&  4&  8& 76&  0& 22&  2\\
        & O & 72& 87& 30& 56& 26& 85& 94& 68& 79& 24& 53& 87& 22& 40& 17& 69& 84& 33& 44& 12& 66& 87& 32& 53& 10& 60& 85& 36& 45& 15& 67& 89& 42& 53& 10& 64& 88& 40& 62& 43& 61& 89& 48& 60& 29\\
        & D &\cellcolor{black!20}{41}&\cellcolor{black!20}{5}&\cellcolor{black!20}{27}&\cellcolor{black!20}{23}&\cellcolor{black!20}{26}&\cellcolor{black!20}{61}&\cellcolor{black!20}{13}&\cellcolor{black!20}{68}&\cellcolor{black!20}{48}&\cellcolor{black!20}{24}&\cellcolor{black!20}{22}&\cellcolor{black!20}{5}&\cellcolor{black!20}{20}&\cellcolor{black!20}{11}&\cellcolor{black!20}{17}&\cellcolor{black!20}{29}&\cellcolor{black!20}{5}&\cellcolor{black!20}{22}&\cellcolor{black!20}{11}&\cellcolor{black!20}{12}&\cellcolor{black!20}{31}&\cellcolor{black!20}{4}&\cellcolor{black!20}{20}&\cellcolor{black!20}{16}&\cellcolor{black!20}{6}&\cellcolor{black!20}{47}&\cellcolor{black!20}{7}&\cellcolor{black!20}{37}&\cellcolor{black!20}{28}&\cellcolor{black!20}{15}&\cellcolor{black!20}{19}&\cellcolor{black!20}{4}&\cellcolor{black!20}{50}&\cellcolor{black!20}{11}&\cellcolor{black!20}{10}&\cellcolor{black!20}{36}&\cellcolor{black!20}{6}&\cellcolor{black!20}{34}&\cellcolor{black!20}{18}&\cellcolor{black!20}{39}&\cellcolor{black!20}{54}&\cellcolor{black!20}{13}&\cellcolor{black!20}{48}&\cellcolor{black!20}{38}&\cellcolor{black!20}{27}\\ \hline
        \multirow{3}{*}{WCA\_NM} & I & 31& 82&  3& 34&  0& 24& 80& -1& 31&  0& 31& 82&  2& 29&  0& 39& 80& 10& 33&  0& 28& 82&  8& 36&  0& 14& 78& -1& 19&  0& 49& 85& -7& 42&  0& 23& 81&  1& 40&  2&  8& 76&  0& 22&  2\\
        & O & 72& 87& 30& 56& 26& 85& 94& 68& 79& 24& 53& 87& 22& 40& 17& 72& 84& 20& 45&  8& 62& 88& 34& 53& 25& 60& 85& 31& 41&  9& 67& 89& 42& 53& 10& 45& 86& 28& 53& 14& 61& 89& 48& 60& 29\\
        & D &\cellcolor{black!20}{41}&\cellcolor{black!20}{5}&\cellcolor{black!20}{27}&\cellcolor{black!20}{23}&\cellcolor{black!20}{26}&\cellcolor{black!20}{61}&\cellcolor{black!20}{13}&\cellcolor{black!20}{68}&\cellcolor{black!20}{48}&\cellcolor{black!20}{24}&\cellcolor{black!20}{22}&\cellcolor{black!20}{5}&\cellcolor{black!20}{20}&\cellcolor{black!20}{11}&\cellcolor{black!20}{17}&\cellcolor{black!20}{33}&\cellcolor{black!20}{5}&\cellcolor{black!20}{10}&\cellcolor{black!20}{11}&\cellcolor{black!20}{8}&\cellcolor{black!20}{33}&\cellcolor{black!20}{7}&\cellcolor{black!20}{26}&\cellcolor{black!20}{17}&\cellcolor{black!20}{25}&\cellcolor{black!20}{45}&\cellcolor{black!20}{6}&\cellcolor{black!20}{31}&\cellcolor{black!20}{22}&\cellcolor{black!20}{9}&\cellcolor{black!20}{19}&\cellcolor{black!20}{4}&\cellcolor{black!20}{50}&\cellcolor{black!20}{11}&\cellcolor{black!20}{10}&\cellcolor{black!20}{22}&\cellcolor{black!20}{5}&\cellcolor{black!20}{27}&\cellcolor{black!20}{13}&\cellcolor{black!20}{12}&\cellcolor{black!20}{54}&\cellcolor{black!20}{13}&\cellcolor{black!20}{48}&\cellcolor{black!20}{38}&\cellcolor{black!20}{27}\\ \hline

    \end{tabular}
    }
\end{table*}
}

\subsection{Experimental Setup}\label{sec:expset}

\subsubsection{SAR Tool Selection}
\mj{To evaluate the effectiveness of \ourtool, we selected two parts of SAR tools. Firstly, we aim to cover as many as latest best-performing SAR tools. To this end, we tried our best to include as many as latest SAR tools published in the recent 5 years. For this part, we included} 42 papers related to architecture recovery technique in the recent five years~\cite{cho_software_2019, huang_multi-agent_2017,link_value_2019, sun_pso_2018, olsson_incremental_2022, izadkhah_information_2019, prajapati_software_2021, hwa_search-based_2017, tarchetti_dct_2020, tan_e-sc4r_2021, elyasi_hygar_2022, papachristou_software_2019, lutellier_measuring_2018, shatnawi_recovering_2017, zahid_evolution_2017, bi_systematic_2018, hoff_towards_2021, teymourian_fast_2022, garcia_constructing_2021, link_recover_2019, link_study_2021, schmitt_laser_arcade_2020, lee_identifying_2020, yang_enhancing_2022, boerstra_stronger_2022, kargar_multi-programming_2019, mohammadi_new_2019, psarras_mechanism_2019, amarjeet_many-objective_2018, rathee_improving_2018, ieva_discovering_2018, benkoczi_design_2018, singh_software_2017, lee_class_2017, li_framework_2017, kargar_semantic-based_2017, marian_hierarchical_2017, sadat2019multi, yano2020moderate, ibrahim2023context, wang2023microservice}. 6 out of the 42 papers~\cite{teymourian_fast_2022, yang_enhancing_2022, boerstra_stronger_2022, psarras_mechanism_2019, papachristou_software_2019,zhang2023software} includes link to their supplementary materials, and 5 of them~\cite{teymourian_fast_2022, yang_enhancing_2022, psarras_mechanism_2019, papachristou_software_2019,zhang2023software} provide artifact for reproduction. Among them, EVOL~\cite{yang_enhancing_2022} and CodeSum~\cite{psarras_mechanism_2019} take some intermediate data as input.
However, we were unable to generate such data for them due to undocumented data formats. Thus, the second part of our SAR tool selection includes the remaining 3 tools: FCA~\cite{teymourian_fast_2022}, SADE~\cite{papachristou_software_2019} and SARIF~\cite{zhang2023software}.
\mj{In addition to the latest tools, in the second part, we aim cover some representative aged SAR tools to ensure the generalizability of \ourtool. To this end, we included} 5 tools that show the most promising results in the previous empirical studies~\cite{garcia2013comparative, lutellier2015comparing, lutellier2017measuring}: ACDC, Bunch, ARC, WCA and LIMBO. 
Among the 8 tools, SADE and SARIF integrate both textual and dependency information for recovery; ARC relies solely on textual information, while the remaining 5 tools are dependency-based.

\subsubsection{SAR Tool Implementation}
The executables of ACDC, Bunch, FCA, SADE, and SARIF are obtained from the author's websites. As for WCA, LIMBO and ARC, the implementations are adopted from ARCADE~\cite{schmitt2020arcade}. Among the 8 tools, Bunch and WCA each have two variants. So a total of 10 implementations will be included in the following evaluation. For both the SAR tools that require external input of structural dependencies and \ourtool, we use DEPENDS~\cite{multilang-depends} as the dependency extractor. For tools
\mj{requiring a preset number of clusters}, we set it to 50. For ARC, the concerns count is set to 100.

\subsubsection{Data Collection}\label{subsubsec_data_collection}


To evaluate the accuracy of a recovered architecture, a commonly-adopted method is comparing it with a human-labeled ground-truth architecture and then evaluating the accuracy based on their similarity. 
We tried our best to gather all projects previously labeled with a ground-truth architecture to form our dataset. We collected a total of 9 open-source projects along with their ground-truth architectures from~\cite{garcia2013obtaining, lutellier2015comparing,zhang2023software}:
ArchStudio4, Bash-4.2, Chromium, Harmony Distributed Camera (HDC), Harmony Drivers Framework (HDF), Hadoop, ITK, Libxml2, and OODT. The details of these projects are provided in Table~\ref{tab_projects}. The size of these projects, measured by lines of code, ranges from 25.7K to 11.7M, which either matches or surpasses the span in prior studies.



\subsubsection{LLM Setup}\label{sec-llm-setup}
\ourtool is implemented with OpenAI's GPT-4o-mini model, which allows a maximum context window of 128k tokens. \mj{We selected GPT-4o-mini as the default model to achieve a balance between the cost and performance. The performance of \ourtool when using other LLMs will be evaluated in RQ4.} 

\mj{More specifically, we used gpt-4o-mini-2024-07-18 model.}
The parameters are kept at the default values, except for the temperature parameter, which is adjusted to 0 to reduce the randomness of GPT's output. \mj{For all conversations, the seed is set to 1 to remove randomness and ensure the reproducibility.}

\subsubsection{Similarities Metrics}\label{subsubsec-sim-metric}

To assess the similarity between a recovered architecture and the ground-truth one, we adopted the 5 metrics following the recent study~\cite{zhang2023software}: MoJoFM~\cite{wen2004effectiveness}, $a2a$~\cite{le2015empirical}, \ccg~\cite{le2015empirical}, ARI~\cite{hubert1985comparing} and \aaj~\cite{zhang2023software}.

MoJoFM~\cite{wen2004effectiveness} measures similarity by the edit distance between two architectures, calculated as the minimum number of operations (moving files between modules or merging modules) needed to transform one architecture into another.
$a2a$~\cite{le2015empirical} is also a distance-based metric that measures the edit distance between architectures, normalized by the total number of operations in both architectures.
$c2c_{cvg}$~\cite{le2015empirical} measures the number of similar clusters between two architectures. We follow~\cite{zhang2023software} to set the similarity threshold to 0.66 in our study.
Adjusted Rand Index (ARI)~\cite{hubert1985comparing} is a well-known metric that quantifies the similarity between two partitions. Unlike the other metrics which range from 0 to 1, ARI ranges from -1 to 1.
$a2a_{adj}$~\cite{zhang2023software} is designed to address the limited variation in the $a2a$ measure by separately considering movement costs and addition/removal costs. Detailed equations for these metrics are available on our website~\cite{oursite}.

\subsection{RQ1: Effectiveness Analysis}\label{subsec-rq1}
In this RQ, we aim to evaluate the effectiveness of \ourtool. Firstly, the effectiveness is evaluated based on the metric improvement after our refinement. Secondly, a case study on Bash is carried out to better understand our results.

\subsubsection{Metric Improvements}

We executed all 10 SAR tools on the 9 projects following the setup detailed in Sec.~\ref{sec:expset}, generating a total of 90 raw architecture recovery results. These results were then refined using \ourtool. 
\mj{As \ourtool generates a hierarchical architecture, it cannot be directly compared to the ground truth architecture using these similarity metrics. Therefore, to make it comparable, we first break the hierarchy that best matches the granularity of ground-truth. Specifically, we select the depth at which the hierarchy has the most similar number of modules to the ground truth. Then we take this partition as a flat representation of our result to compare with the input ones. Table~\ref{tab_metrics_result} presents the metric scores for the input/refined architectures, and the improvements measured in percentage points (PP.), showing the absolute score differences. }

\begin{table}[tb]
    \centering
    \small
    \caption{Averaged Metric Improvements}
    \label{table-avg-score}
    \begin{tabular}{l|ccccc}
        \hline
            & MoJoFM       & a2a & ARI & \aaj      & \ccg\\ \hline\hline
        IN & 0.3935    & 0.8142   & 0.1127  & 0.3355 & 0.1151      \\
        OUT & 0.6684    & 0.8784   & 0.3800  & 0.5434 & 0.2775      \\\hline
        PP. & 0.2749    & 0.0642   & 0.2672  & 0.2078 & 0.1687      \\
        INC & 118.57\%    & 8.06\%  & NaN & 100.41\% & NaN     \\
        RDP & 43.35\%     & 33.03\%   & 29.39\%  & 30.18\%  & 17.72\%     \\ \hline
        \end{tabular}
\end{table}

From Table~\ref{tab_metrics_result}, we observe that for all 90 input architectures, \ourtool consistently improves accuracy across all five metrics. This result demonstrates the robustness of \ourtool. We also present the average effect of \ourtool in Table~\ref{table-avg-score}. Specifically, we report the average absolute percentage point (PP) improvement and the percentage increment compared to the input (INC). The percentage increment for ARI and \ccg is omitted because some input scores are close to zero, leading to disproportionately high and meaningless percentage increments.


Based on the PP and INC values, we find that MoJoFM, ARI, and \aaj show significant improvements after applying \ourtool, with absolute increments ranging from 0.2078 to 0.2775, representing more than a 100\% increase in metric scores. Moreover, \ccg also achieves a notable PP improvement of 0.1599. However, both PP and INC for a2a remain limited. This is due to the a2a metric’s restricted dynamic range~\cite{lutellier2015comparing}, as it typically scores above 0.7 even when the architecture quality is poor. This characteristic results in lower PP and INC values for a2a.

\begin{figure}[tb]
    \centering
    \includegraphics[width=0.48\textwidth]{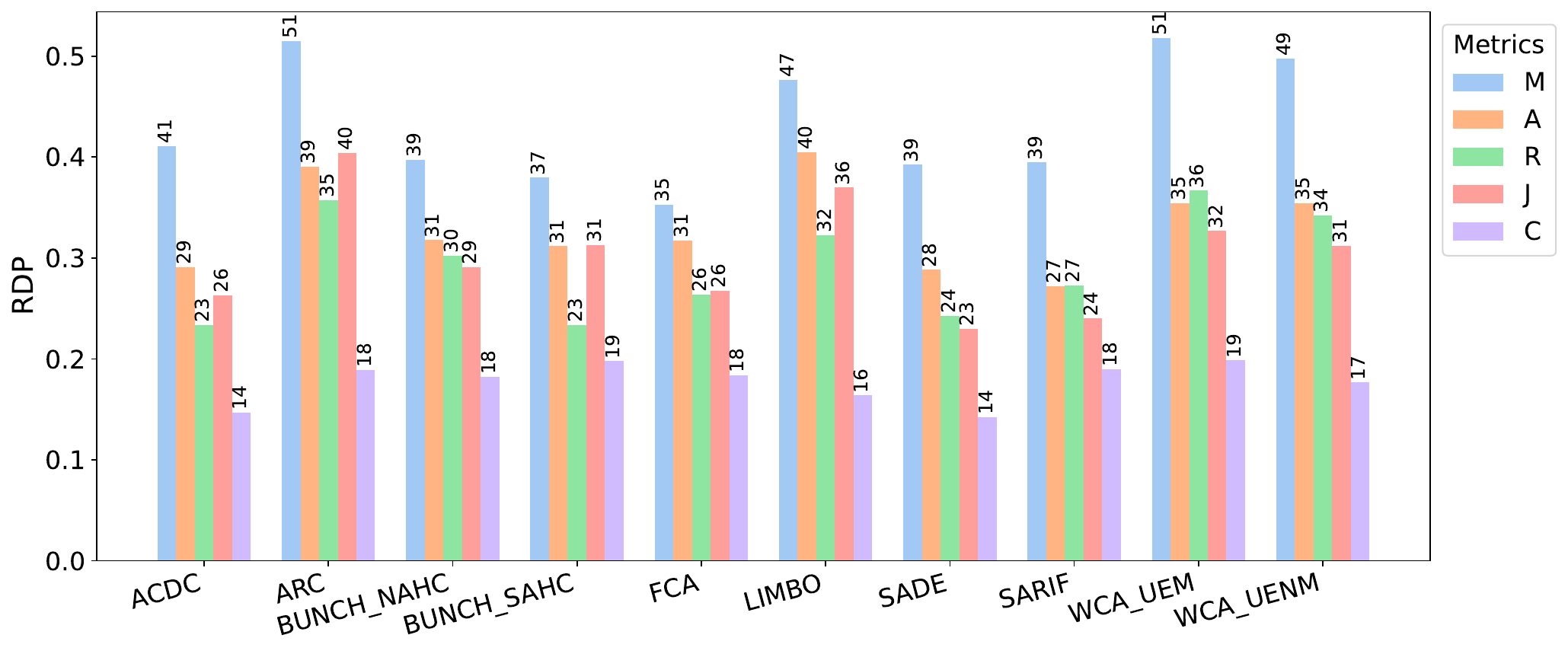}
    \caption{$RDP$ averaged over tools}
    \label{fig-tool-rdp}
\end{figure}

Besides the impact of the limited range of a2a, we argue that using PP or INC to measure improvement has inherent limitations. In many cases, increasing a score from 0.5 to 0.6 is much less meaningful than improving it from 0.9 to 1.0, even though both have an absolute difference of 0.1. To address this issue, we introduce Reduction in Distance to Perfect (RDP), also known as normalized gain~\cite{bao2006theoretical}, to measure improvement. RDP is defined as:

\begin{equation}\label{eq-rdp}
    RDP(A, A') = \frac{M(A')-M(A)}{1-M(A)}
\end{equation}
where $A$ and $A'$ are the input and output architecture, $M(A)$ is the metric value of $A$. 
As its name suggests, RDP measures the percentage reduction in the distance to the perfect score of the metric (i.e., 1). In our context, the RDP value indicates the percentage of errors in the input architecture that have been eliminated after refinement, aligning well with \ourtool's objective.
As shown in Table~\ref{table-avg-score}, when measured by RDP, MoJoFM, a2a, ARI, and \aaj all exhibit significant improvements, with increases ranging from 28.18\% to 43.32\%. The \ccg metric also shows a notable improvement, with an increment of 16.79\%.
Specifically, by combining the equation for RDP (Eq.~\ref{eq-rdp}) with the equation for MoJoFM (See~\cite{wen2004effectiveness}), we derive:

\begin{equation}\label{eq-mojo3}
    RDP(A, A') = 1-\frac{mno(A',GT)}{mno(A,GT)}
\end{equation}

As introduced in Sec.~\ref{subsubsec-sim-metric}, $mno(A,GT)$ 
represents the number of operations required to transform architecture $A$ into the ground truth. Therefore, the RDP of MoJoFM indicates the percentage reduction in the number of operations needed to align the architecture with the ground truth after applying \ourtool. As shown in Table\ref{table-avg-score}, this value is 43.35\%.

\mybox{Finding 1: \ourtool improved the accuracy of all 90 input architectures, regardless of the metric used. The absolute increments across the five metrics range from 0.0627 to 0.2749. When measured by RDP, the improvements range from 17.72\% to 43.35\%. Specifically, based on MoJoFM, the edit distance to the ground truth architecture was reduced by 43.35\% after applying \ourtool.}

\begin{figure}[tb]
    \centering
    \includegraphics[width=0.48\textwidth]{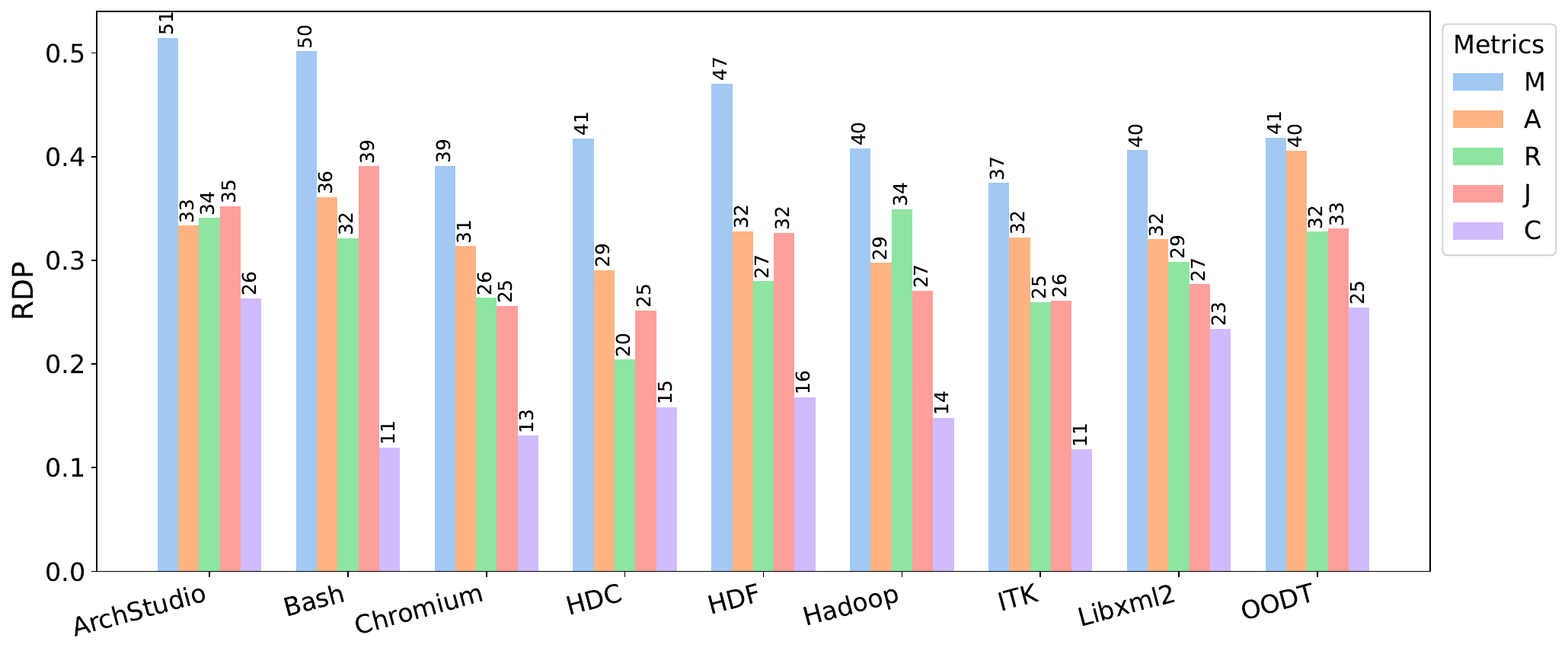}
    \caption{$RDP$ averaged over projects}
    \label{fig-prj-rdp}
\end{figure}

\begin{table}[tb]
    \centering
    \small
    \caption{P-values of ANOVA test}
    \label{table-anova}
    \begin{tabular}{l|ccccc}
        \hline
            & MoJoFM       & a2a & ARI & \aaj      & \ccg\\ \hline
        Tool & 0.0622    & 0.2793   & 0.2646  & 0.0627 & 0.9541      \\
        Project & 0.2566    & 0.5853   & 0.3369  & 0.1227 & 7.50E-5      \\\hline
        \end{tabular}
\end{table}

Next, we evaluate how the performance of \ourtool is affected by the input SAR tool and the project. To do this, we measure the average RDP for each tool and project, with the results shown in Fig.~\ref{fig-tool-rdp} and Fig.~\ref{fig-prj-rdp}.
As the figures show, for most metrics, the improvement measured by RDP does not significantly vary across different input tools or projects. To statistically test whether \ourtool's performance is influenced by either the tool or the project, we conducted a series of one-way analyses of variance (ANOVA)~\cite{fisher1970statistical}. The resulting p-values for each metric are shown in Table~\ref{table-anova}. For tools, none of the five metrics show significant differences across different tools, as all p-values are greater than 0.05. For projects, four of the metrics also show no significant differences. However, the improvement in \ccg is affected by the project, with a p-value much smaller than 0.05.
From Fig.~\ref{fig-prj-rdp}, we can observe that for ArchStudio, Libxml2, and OODT, the increase in \ccg is much larger than for the other projects. This could be because these projects have many small modules in their ground truth. As previously mentioned, \ccg is calculated based on the degree of overlap between ground-truth modules and recovered modules. The presence of smaller ground-truth modules makes such identification easier, leading to differences in \ccg across projects.

\begin{figure*}[t]
    \centering
    \centering
    \includegraphics[width=0.8\textwidth]{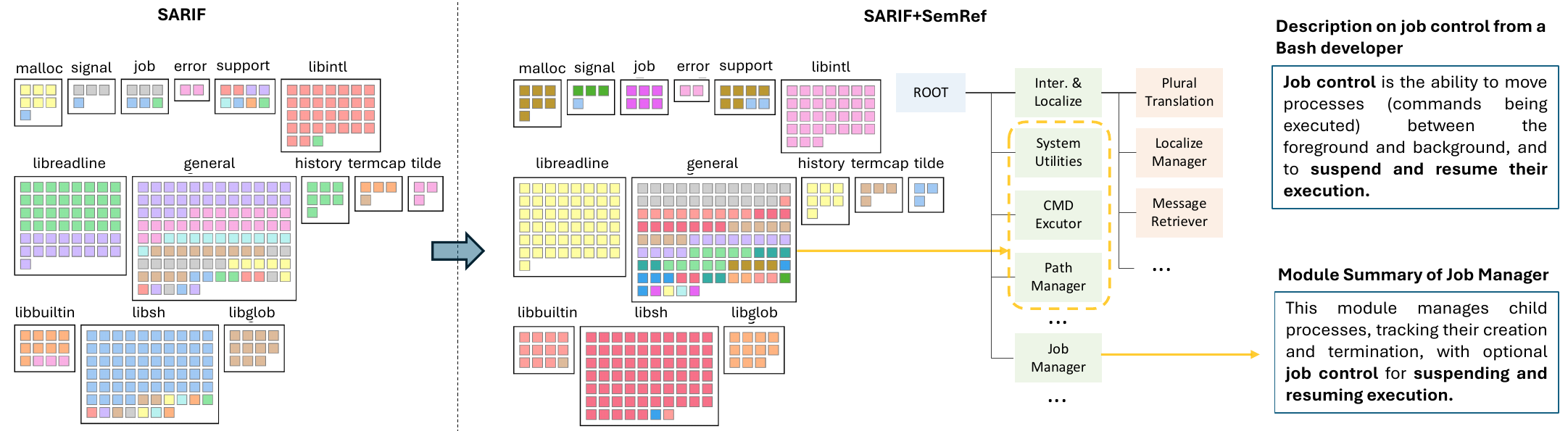} 
    \caption{Case study on Bash}
    \label{fig-bash}
\end{figure*}

\mybox{Finding 2: When measured by RDP, the effectiveness of \ourtool on \ccg can vary across different projects. In all other cases, \ourtool's effectiveness, as measured by RDP, shows no significant difference across either input tools or projects.}


{
\setlength{\tabcolsep}{1.85pt}
\begin{table}[tb]
    \centering
    \scriptsize
    \caption{\mj{Metrics based on top and lowest level modules (D: default strategy, T: top-level modules, L: leaf modules)}}
    \label{table-avg-score-2}    
    \mj{
    \begin{tabular}{l|ccc|ccc|ccc|ccc|ccc}
    \hline
    \multirow{2}{*}{Tools} & \multicolumn{3}{c|}{MoJoFM} & \multicolumn{3}{c|}{a2a}  & \multicolumn{3}{c|}{ARI}  & \multicolumn{3}{c|}{\aaj} & \multicolumn{3}{c}{\ccg} \\
                        & D       & T       & L       & D       & T      & L      & D       & T      & L      & D            & T           & L           & D           & T           & L           \\ \hline
    IN                     & \multicolumn{3}{c|}{.394}   & \multicolumn{3}{c|}{.814} & \multicolumn{3}{c|}{.113} & \multicolumn{3}{c|}{.335}                & \multicolumn{3}{c}{.115}                \\
    OUT                    & .668    & .565    & .630    & .878    & .863   & .856   & .380    & .307   & .308   & .543         & .483        & .485        & .278        & .247        & .235        \\ \hline
    PP.                    & .275    & .172    & .237    & .064    & .049   & .042   & .267    & .194   & .195   & .208         & .148        & .150        & .169        & .129        & .117        \\
    INC(\%)                & 118     & 81.8    & 108     & 8.06    & 6.24   & 5.39   & NaN     & NaN    & NaN    & 100          & 80.7        & 78.2        & NaN         & NaN         & NaN         \\
    RDP(\%)                & 43.4    & 24.0    & 35.1    & 33.0    & 24.4   & 19.6   & 29.4    & 20.5   & 20.3   & 30.2         & 20.4        & 20.5        & 17.7        & 13.2        & 11.8        \\ \hline
    \end{tabular}
    }
\end{table}
}
As a refinement framework, \ourtool can also function as a SAR tool when integrated with an existing SAR tool. Since \ourtool shows no significant preference for input tools, we selected the most promising SAR tool, SARIF, to combine with \ourtool and form a new SAR tool. We then compared the accuracy of SARIF+\ourtool against the best baseline result, as shown in Table~\ref{table-combine}.
The best baseline score refers to the highest score produced by any of the 10 SAR tools for a given project and metric. According to the results, the combination of \ourtool and SARIF outperforms all other SAR tools across the 9 projects. When measured by RDP, our approach was 24.49\% closer to the ground truth across five metrics on average.

\mybox{Finding 3: Combined with SARIF, \ourtool produces the most promising architecture recovery results across all 9 projects. Compared with the best baselines, our result is 24.49\% better than the best baselines.}


\mj{
As stated previously, our reported metrics were based on the hierarchical level that has the closest number of clusters to the ground-truth. This choice could raise a concern that the observed improvement was primarily due to this fattening strategy.
To address this concern, we evaluated two additional strategies: 1) using only the top-level modules, and 2) using only the lowest-level (leaf) modules. Table~\ref{table-avg-score-2} shows the averaged metrics for these strategies.
As shown in the table, \ourtool still achieves significant accuracy improvements when using either the top-level or the leaf modules. This is true even though the granularity of these modules may differ greatly from the ground truth. Specifically, when measured by RDP, accuracy improved by 13.2\% to 24.0\% for top-level modules. For leaf modules, the accuracy improved by 11.8\% to 35.1\%.}

\mybox{\mj{Finding 4: \ourtool can increase the accuracy by 13.2\% to 24.0\% and 11.8\% to 35.1\% when taking top-level modules or leaf modules as results (measured by RDP).}}



{
\setlength{\tabcolsep}{2.85pt}
\begin{table}[tb]
    \centering
    \caption{Accuracy of SARIF+\ourtool against best baseline} \footnotesize
    \label{table-combine}
    \scalebox{0.9}{
    \begin{tabular}{l|ccccc|ccccc|ccccc}
    \hline
    & \multicolumn{5}{c|}{ArchStudio} & \multicolumn{5}{c|}{Bash} & \multicolumn{5}{c}{Chromium}\\
    & {\scriptsize M} & {\scriptsize A} & {\scriptsize R} & {\scriptsize J} & {\scriptsize C} & {\scriptsize M} & {\scriptsize A} & {\scriptsize R} & {\scriptsize J} & {\scriptsize C} & {\scriptsize M} & {\scriptsize A} & {\scriptsize R} & {\scriptsize J} & {\scriptsize C}\\\hline
    Best Baseline &73&87&33&51&42  &76&90&51&67&17  &63&87&32&49&24\\
    SARIF+\ourtool &84&92&63&70&46  &82&91&62&72&26  &79&90&47&57&24\\
    PP. &\cellcolor{black!20}{11}&\cellcolor{black!20}{5}&\cellcolor{black!20}{29}&\cellcolor{black!20}{19}&\cellcolor{black!20}{4}    &\cellcolor{black!20}{6}&\cellcolor{black!20}{1}&\cellcolor{black!20}{11}&\cellcolor{black!20}{5}&\cellcolor{black!20}{9}    &\cellcolor{black!20}{16}&\cellcolor{black!20}{2}&\cellcolor{black!20}{15}&\cellcolor{black!20}{7}&\cellcolor{black!20}{0}\\\hline\hline
    
    & \multicolumn{5}{c|}{HDC} & \multicolumn{5}{c|}{HDF} & \multicolumn{5}{c}{Hadoop}\\
    & {\scriptsize M} & {\scriptsize A} & {\scriptsize R} & {\scriptsize J} & {\scriptsize C} & {\scriptsize M} & {\scriptsize A} & {\scriptsize R} & {\scriptsize J} & {\scriptsize C} & {\scriptsize M} & {\scriptsize A} & {\scriptsize R} & {\scriptsize J} & {\scriptsize C}\\\hline
    Best Baseline &89&88&51&54&16  &59&88&35&50&22  &54&87&52&56&23\\
    SARIF+\ourtool &92&92&62&63&33  &68&90&44&58&29  &69&92&70&66&41\\
    PP. &\cellcolor{black!20}{3}&\cellcolor{black!20}{3}&\cellcolor{black!20}{11}&\cellcolor{black!20}{10}&\cellcolor{black!20}{18}    &\cellcolor{black!20}{8}&\cellcolor{black!20}{2}&\cellcolor{black!20}{8}&\cellcolor{black!20}{8}&\cellcolor{black!20}{6}    &\cellcolor{black!20}{14}&\cellcolor{black!20}{4}&\cellcolor{black!20}{18}&\cellcolor{black!20}{10}&\cellcolor{black!20}{18}\\\hline\hline

    & \multicolumn{5}{c|}{ITK} & \multicolumn{5}{c|}{Libxml2} & \multicolumn{5}{c}{OODT}\\
    & {\scriptsize M} & {\scriptsize A} & {\scriptsize R} & {\scriptsize J} & {\scriptsize C} & {\scriptsize M} & {\scriptsize A} & {\scriptsize R} & {\scriptsize J} & {\scriptsize C} & {\scriptsize M} & {\scriptsize A} & {\scriptsize R} & {\scriptsize J} & {\scriptsize C}\\\hline
    Best Baseline &76&85&21&42&0  &53&90&38&58&25  &50&86&24&45&17\\
    SARIF+\ourtool &89&88&42&49&19  &74&92&46&69&41  &61&89&53&63&51\\
    PP. &\cellcolor{black!20}{13}&\cellcolor{black!20}{2}&\cellcolor{black!20}{21}&\cellcolor{black!20}{7}&\cellcolor{black!20}{19}    &\cellcolor{black!20}{20}&\cellcolor{black!20}{2}&\cellcolor{black!20}{8}&\cellcolor{black!20}{10}&\cellcolor{black!20}{16}    &\cellcolor{black!20}{11}&\cellcolor{black!20}{3}&\cellcolor{black!20}{29}&\cellcolor{black!20}{17}&\cellcolor{black!20}{33}\\\hline
    \end{tabular}
    }
\end{table}
}

\subsubsection{Case Study on Bash}

To further validate \ourtool, we conducted a case study on Bash with the architecture recovered by SARIF. The architectural recovered by SARIF and our refinement result are shown in Fig.~\ref{fig-bash}.
Both the original and refined architectures are presented by a comparison with the ground truth architecture. In the comparison figure, small colored squares represent files, the color-coding indicates the file grouping in the recovered architecture, and the boxes show the ground truth architecture.

From the figure, we can observe that the refined architecture are much more closer to the ground truth one. For example, for \textit{libsh}, although most of the files were initially grouped together in SARIF's result, many were still scattered across other modules. After the refinement, most of these files were correctly regrouped into the \textit{libsh} module.
For large modules in the ground truth such as \textit{libintl}, compared to the input architecture, we not only corrected its file grouping but also provided a more detailed view by identifying its submodules.
Moreover, for modules like \text{job}, its six files were scattered across three modules in SARIF's results. \ourtool not only accurately identified this module but also provided a name (\textit{Job Manager}) and a summary. As shown in the figure, we compared \ourtool's summary with the developer's description of Bash's job control~\cite{bash-book} and found them to be semantically similar. Such a name and summary can help users quickly understand a project, especially in the absence of documentation.

Most of the modules in our refined result closely match the ground truth. Yet the \textit{general} module is an exception, which is divided into multiple parts in our result. Although there is a utility module in our result, some modules, such as \textit{Command Executor} and \textit{Path Manager}, are not grouped within it. We assert that our division is also reasonable, as these modules serve distinct and important functions for \textit{Bash}.




\subsection{RQ2: Ablation Study}
\begin{figure}[t]
    \centering
    \centering
    \includegraphics[width=0.45\textwidth]{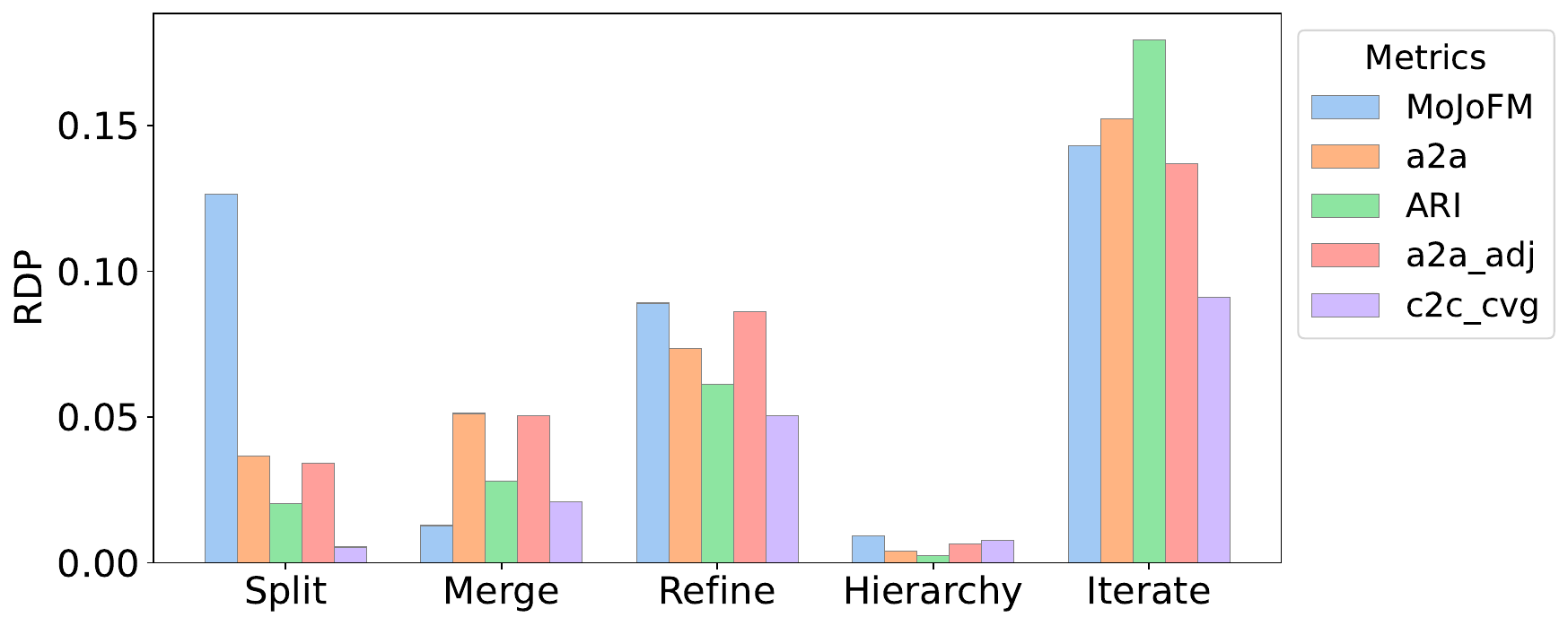} 
    \caption{Acc. improvement of each stage measured by $RDP$}
    \label{fig-ablation-red}
\end{figure}

In this RQ, we aim to explore how each part of our design contributed to the final result. To this end, we output the intermediate refinement result for each stage by removing the stages after it. The stages tested include: 1) splitting, 2) merging, 3) file-level refinement, 4) hierarchy construction, and 5) iterative refinement. The RDP value compared with the output of the previous stage is reported in Fig.~\ref{fig-ablation-red}.

As shown in Fig.~\ref{fig-ablation-red}, only the hierarchy construction step has a low impact on the accuracy. This result is intuitive since its primary purpose is to organize flat modules into a hierarchy rather than refining them. For the remaining four stages, the iterative refinement made the largest contribution to the accuracy no matter measured by which metrics. However, for splitting, merging, and file-level refinement, their contribution differs for different metrics. 


For a2a, ARI, \aaj, and \ccg, the contributions to accuracy follow this order: file-level refinement > merging > splitting. However, this pattern does not hold for MoJoFM. For MoJoFM, splitting provides the largest improvement, while merging has almost no effect on the metric.
This is because, in MoJoFM, the cost of merging two modules is extremely low, while the cost of splitting them is significantly higher. As a result, even when a module is correctly merged into another, the reduction in cost from the current architecture to the ground truth is minimal. In contrast, correctly splitting a module leads to a substantial cost reduction, resulting in a significant improvement in the MoJoFM score.

\mybox{Finding \mj{5}: The Iterative Refinement stage contributes the most to accuracy. The file-level refinement, merging, and splitting stages also provide a significant contribution. While the hierarchy construction stage has only a minor impact on accuracy, it serves as a crucial foundation for the subsequent stages.}

\subsection{RQ3: Cost Analysis}
\begin{figure}[t]
    \centering
    \subfigure[\#Token v.s. project size]{
        \includegraphics[width=0.19\textwidth]{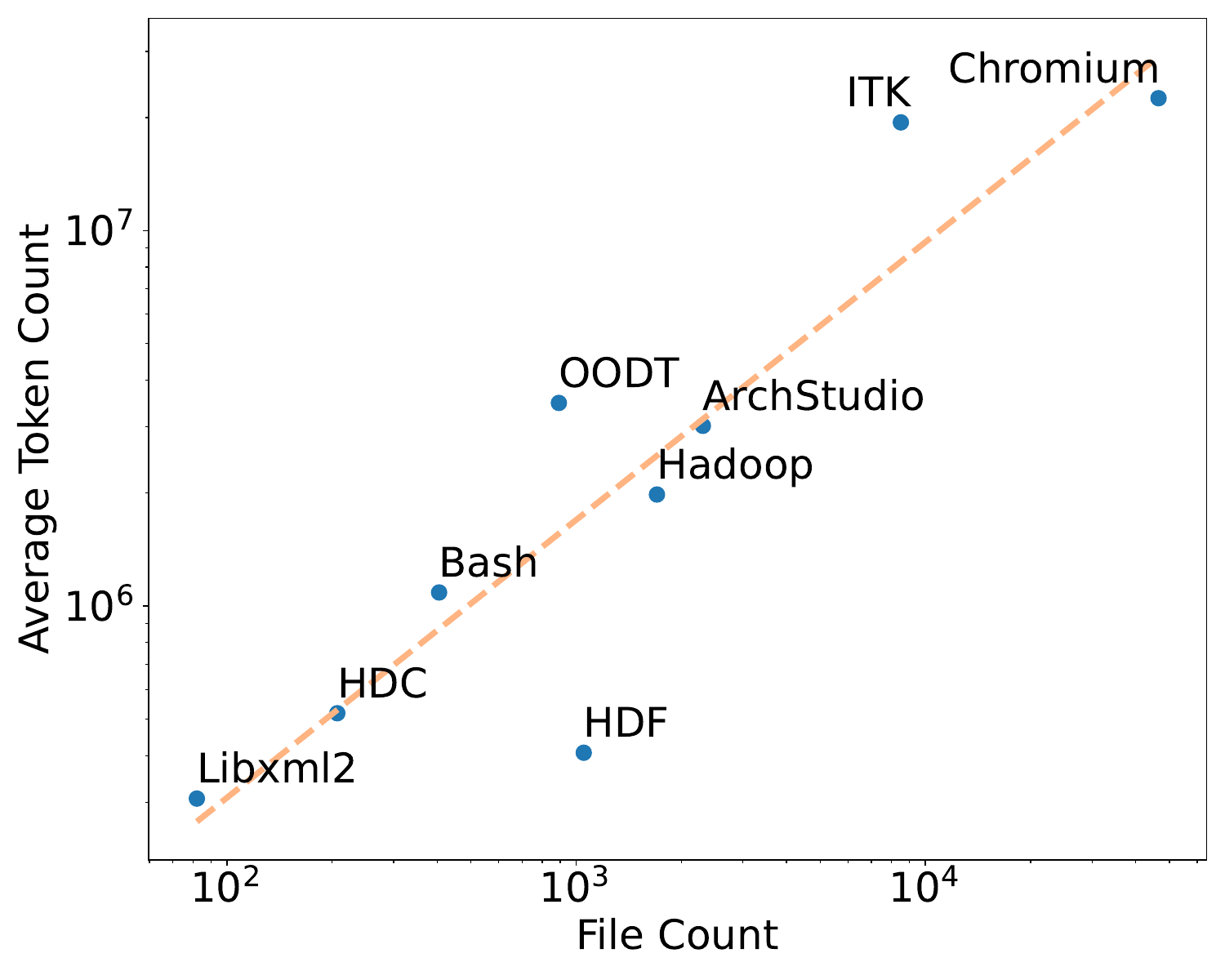}
        \label{subfig-token-cnt}
    }
    \subfigure[Usage breakdown]{
        \includegraphics[width=0.23\textwidth]{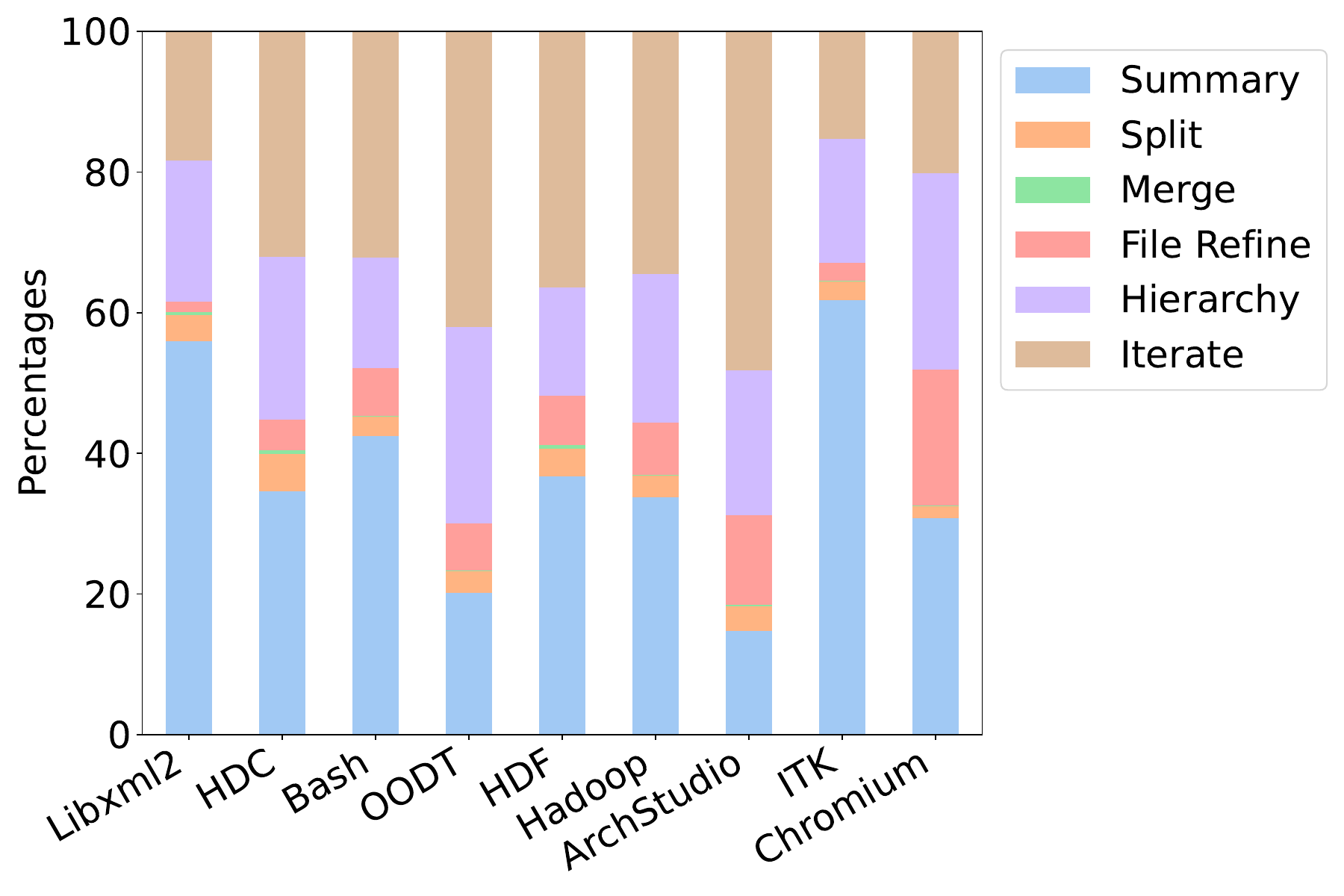}
        \label{subfig-token-proportion}
    }
    \caption{Statistic of token usage}
    \label{fig-cost}
\end{figure}

In this RQ, we aim to explore how many tokens are required for refining an architecture and how it scales with the project size.
We first analyzed for each project, how many tokens are consumed on average by \ourtool. The results are shown in Fig.~\ref{subfig-token-cnt}. As depicted in the figure, the average number of tokens used for refining each project is approximately linearly proportional to the project size measured by file count. For Chromium, the largest project in our dataset, a total of 18.16M tokens were used for \ourtool to perform the refinement. Based on the pricing from OpenAI at Jan 2025, it would cost 2.72 USD to call the API from the GPT-4o-mini model. 

Next, we broke down the token cost for each step to gain a more detailed understanding. For each project, the average percentage of tokens spent on each step is shown in Fig.~\ref{subfig-token-proportion}. Although the percentage varies among the projects, the overall trend is similar: summarizing files and building the hierarchy are the most costly steps, while the preprocessing steps account for only a very small part of the total cost.

\mybox{Finding \mj{6}: The total token count increases approximately linearly with the project size, as measured by the number of files.}


\subsection{RQ4: Model Comparison}
\mj{
In the previous RQs, we investigated \ourtool's performance based on GPT-4o-mini. However, as a refinement framework, \ourtool should work with any underlying LLMs. Therefore, in this RQ, we aim to verify the generalizability of \ourtool across various LLMs. To this end, we include 4 alternative LLMs. Two of them are powerful commercial LLMs: one is GPT-4o (gpt-4o-2024-08-06), which is a direct superior model to GPT-4o-mini by OpenAI; another is Gemini-2.0-flash-001, which is one of the latest stable LLM by Google. Besides the commercial ones, we also included two open-source LLMs: Llama 3.3 (70b)~\cite{grattafiori2024llama} and DeepSeek-V3-0324. 
}

\mj{
    We then tested 4 LLMs on the SARIF results on all 9 projects. We show their averaged improvements as measured by RDP in Table~\ref{table-models-avg} (detailed results are available on our website~\cite{oursite}). Three models completed the task successfully, while Llama3.3 failed on the two largest projects (ITK and Chromium). The main reason for this failure is that Llama3.3 sometimes did not follow the required JSON schema, especially when processing long inputs. Moreover, it still performed worse overall, considering only the successful projects. For the other three LLMs, GPT-4o, Gemini-flash, and DeepSeek-V3 achieved better results than GPT-4o-mini by showing higher average improvements for most metrics.
    
    }

\mybox{\mj{Finding 7: Using \ourtool with GPT-4o, Gemini-flash and DeepSeek-V3 produce better results than using GPT-4o-mini. Llama3.3 (70B) provide worse result than GPT-4o-mini and failed on two large projects.}}


\begin{table}[tb]
    \centering
    \small
    
    \caption{\mj{Averaged RDP across Different LLMs on SARIF input}}
    \label{table-models-avg}
    \mj{
    \begin{tabular}{l|cccccc}
    \hline
            & MoJoFM & a2a    & ARI    & \aaj & \ccg & Fail \\ \hline
    GPT-4o-mini  & 39.5\% & 28.9\% & 27.3\% & 24.1\%              & 19.3\%              & 0       \\
    Llama3.3 (70B)*   & 18.7\% & 8.49\% & 9.36\% & 6.01\%              & 4.63\%              & 2       \\
    GPT-4o   & 41.5\% & 23.6\% & 35.0\% & 29.4\%              & 20.4\%              & 0       \\
    Gemini-2.0-flash   & 41.6\% & 23.4\% & 28.4\% & 27.2\%              & 21.6\%              & 0       \\
    Deepseek-V3 & 44.2\% & 34.3\% & 33.0\% & 25.7\%              & 19.7\%              & 0       \\ \hline
    \end{tabular}
    
    \begin{flushleft}
    \footnotesize
    * Averaged over successful projects.
    \end{flushleft}
    }
\end{table}



\mj{\section{Discussion}}
\mj{
\subsection{Implications}}
\mj{

\noindent\textbf{Improvements by \ourtool.} 
In RQ1, we showed that \ourtool improves accuracy and is stable across various projects and inputs. Besides improving accuracy, \ourtool offers two other benefits. First, it transforms modules with a flat architecture into a hierarchical one. This new structure allows users to focus on different abstraction levels depending on their needs. Second, it generates a summary for each module, which helps users to quickly understand its capability.

\noindent\textbf{Input tool selection.} 
Based on Finding 2 in RQ2, \ourtool's effectiveness (measured by RDP) is similar across different input tools. This means users should choose the best available SAR tool as input for optimal results. Currently, SARIF is the best option. When more powerful tools become available, using them as input should produce better results. 
Companies with proprietary architecture recovery tools can also use them as input to \ourtool.

\noindent\textbf{Model selection.}
In RQ4, we tested four alternative LLMs to understand how model choice affects \ourtool's performance. The smaller Llama3.3 (70B) model showed limitations by failing on two large projects and generally performing worse than other models. In contrast, the remaining models all produced better results than GPT-4o-mini. This performance pattern aligns with their general capabilities as measured by the GPQA benchmark~\cite{rein2023gpqagraduatelevelgoogleproofqa}. Such scaling means that in the future, as more powerful LLMs become available, users can adopt them to achieve better results. This ensures that \ourtool will continue to improve alongside advances in LLM technology.
These results also show that \ourtool works with different LLM providers. It performs well with both commercial and open-source models, which allow users to choose LLMs flexibly. 



}
\mj{\subsection{Threats to Validity}\label{sec_threats}}



\mj{The first threat is the potential for hallucinations by the LLMs. To mitigate this, we employed chain-of-thought prompting throughout our approach. For critical decisions such as adjustments to the module hierarchy, we used self-consistency checks by posing multiple questions and only accepting the LLM's decisions when all answers confirmed them. However, we acknowledge that hallucination can never be fully eliminated. Further prompt optimization, like using few-shot prompting, remains a direction for future work.}

The second threat is potential data leakage. Among the ground truth architectures we collected, Libxml2, HDC, and HDF were obtained from \cite{zhang2023software}, which was published after the knowledge cutoff date of GPT-4o-mini (Oct 2023), while the remaining ground truths were published before this date. According to our results, \ourtool's performance shows no significant difference across projects for four metrics. 
This result suggests that data leakage has a minimal impact on our performance.

The third potential threat is the size of the evaluation dataset. Although we tried our best to collect as many projects with human-labeled ground truth as possible, we managed to gather only 9 projects. Nevertheless, these projects cover a broad spectrum of sizes, programming languages, and domains, which enhances their representativeness. Furthermore, to the best of our knowledge, our evaluation dataset is either larger than or comparable to those used in previous studies.

\section{Conclusion}\label{sec_conclusion}
In summary, we proposed \ourtool, an LLM-based framework for refining SAR results. We evaluated \ourtool on 10 existing SAR tools across 9 projects. The results show that \ourtool improves accuracy by 17.72\% to 43.35\% across different metrics, as measured by $RDP$. The cost of \ourtool scales approximately linearly with project size, confirming its scalability. Furthermore, we tested \ourtool on various LLMs, demonstrating that using more advanced LLMs in the future could further enhance our performance.

\section{Data Availability.}
The data and codes are available on \mj{our website~\cite{oursite} or Zenodo~\cite{zenodo-site}.}

\begin{acks}
This research is part of the IN-CYPHER programme and is supported by the National Research Foundation, Prime Minister’s Office, Singapore under its Campus for Research Excellence and Technological Enterprise (CREATE) programme; by the National Key Research and Development Program of China (Grant No. 2024YFF0908000); by the National Research Foundation, Singapore, and DSO National Laboratories under the AI Singapore Programme (AISG Award No: AISG4-GC-2023-008-1B); and by the National Research Foundation Singapore and the Cyber Security Agency under the National Cybersecurity R\&D Programme (NCRP25-P04-TAICeN).
Any opinions, findings and conclusions, or recommendations expressed in these materials are those of the author(s) and do not reflect the views of the National Research Foundation, Singapore, Cyber Security Agency of Singapore, Singapore.
\end{acks}

\clearpage

\bibliographystyle{ACM-Reference-Format}
\bibliography{ref-arch-rec}

\end{document}